\documentclass[11pt,a4paper]{article}%
\usepackage{jheppub}
\usepackage{amsmath}
\usepackage{amssymb}
\usepackage{amsfonts}
\usepackage{relsize}
\usepackage{enumerate}
\usepackage{booktabs}
\usepackage{xcolor}
\usepackage{float}
\usepackage{placeins}
\usepackage{graphicx}
\usepackage{epstopdf}
\usepackage[justification=justified,format=plain]{caption}
\usepackage{subcaption}
\usepackage{physics}
\usepackage{hyperref}
\usepackage{xspace}
\usepackage{bm}
\usepackage{titlesec}
\usepackage{comment}
\usepackage[skip=10pt plus1pt, indent=0pt]{parskip}
\usepackage[title,titletoc,toc,page]{appendix}
\graphicspath{./image/}
\renewcommand{\theequation}{\arabic{section}.\arabic{equation}}

\DeclareFontFamily{U}{mathb}{\hyphenchar\font45}
\DeclareFontShape{U}{mathb}{m}{n}{
      <5> <6> <7> <8> <9> <10> gen * mathb
      <10.95> mathb10 <12> <14.4> <17.28> <20.74> <24.88> mathb12
      }{}
\DeclareSymbolFont{mathb}{U}{mathb}{m}{n}
\DeclareFontSubstitution{U}{mathb}{m}{n}

\let\dot\relax
\DeclareMathAccent{\dot}{0}{mathb}{"39}
\let\ddot\relax
\DeclareMathAccent{\ddot}{0}{mathb}{"3A}
\let\dddot\relax
\DeclareMathAccent{\dddot}{0}{mathb}{"3B}
\let\ddddot\relax
\DeclareMathAccent{\ddddot}{0}{mathb}{"3C}
\begin{document}

\title{Page Curve of AdS-Vaidya Model for Evaporating Black Holes}

\affiliation[a]{Department of Electrophysics, National Yang Ming Chiao Tung University, Hsinchu, ROC}
\affiliation[b]{Center for Theoretical and Computational Physics, National Yang Ming Chiao Tung University, Hsinchu, ROC}
\emailAdd{chiajuichou@nycu.edu.tw}
\emailAdd{hanslao.sc07@nycu.edu.tw}
\emailAdd{yiyang@nycu.edu.tw}
\author{Chia-Jui Chou${^{a}}$, Hans B. Lao${^{a}}$ Yi Yang${^{ab}}$ }
%%%%%%%%%%%%%%%%%%%%%%%%%%%%%%%%
\abstract{We study an evaporating black hole in the boundary conformal field theory (BCFT) model under the fully time-dependent AdS-Vaidya spacetime geometry. We introduce the time-dependent finite bath termed the effective Hawking radiation region. This is described by a nontrivial BCFT solution that acts as a time-dependent brane which we call the moving end-of-the-radiation (METR) brane that leads to a new type of Hubeny-Rangamani-Takayanagi surface. We further examine the island formulation in this particular time-dependent spacetime. The Page curve is calculated by using Holographic Entanglement Entropy (HEE) in the context of double holography.}

\maketitle

%%%%%%%%%%%%%%%%%%%%%%%%%%%%%%%%
\section{Introduction and Summary}

The predictions of Hawking based on semiclassical effective field theory suggest that black holes emit radiation similar to black bodies with a corresponding temperature. As a consequence of Hawking radiation, a black hole should eventually evaporate away if there is no ingoing matter to compensate for the loss of energy \cite{BF02345020,PhysRevD.13.191}. This phenomenon encapsulates the essence of the information loss paradox for which if we assume that the black hole forms in a pure state, it ends up in a mixed state after evaporation which essentially violates one of the tenets of quantum mechanics, i.e., unitarity principle. The fine-grained entropy of the radiation increases linearly until the black hole evaporates away, while during the increase it exceeds the coarse-grained entropy of the black hole. The fine-grained entropy of the black hole, which is bounded above by the coarse-grained entropy, should decrease during evaporation. Don Page has demonstrated that if a black hole were to evaporate obeying unitarity assuming the system formed in a pure state so that the fine-grained entropy of the radiation and the black hole must be equal, the fine-grained entropy of radiation should follow an increasing and eventually decreasing behavior which is called the Page curve with a Page time at the saturation point \cite{9306083, 1301.4995}.

There were many proposals to try and alleviate the issues posed by the information paradox \cite{1207.3123, 1211.6767, 1301.4504, 1304.6483, 1306.0533, 1308.4209, 1310.6334, 1607.05141, 1712.04955, 1910.00972, 1911.06305, 2002.03543,2103.07477,2105.06579,2105.12737,2109.14618,2110.04233,2111.09551}. Among the proposals, one route that has taken much attention is through the Ryu and Takayanagi (RT) prescription which studies the fine-grained entropy in a geometrical context through the AdS/CFT correspondence, albeit only for time-independent cases \cite{0603001,0605073}. This was generalized further to time-dependent cases by Hubeny, Rangamani, and Takayanagi (HRT) \cite{0705.0016}. The fine-grained entropy or sometimes also called the entanglement entropy computed in this manner is usually called the holographic entanglement entropy (HEE). An even further generalization of the HRT prescription was due to the incorporation of quantum corrections which lead to the generalized entropy in a CFT \cite{1307.2892}, and in the context of holography conducted the proposal of including the bulk entropy and finding a quantum extremal surface (QES) \cite{1408.3203,1904.08423}.

Building on these ideas, a concrete proposal was introduced to resolve the information paradox which is formally called the \textit{island rule} \cite{1905.08255,1905.08762,1908.10996}. The rule is given as follows, one first computes the generalized entropy $S_{\rm{gen}}$ as,
\begin{equation}
  S_{\rm{gen}} (\mathcal{R} \cup \mathcal{I}) = \frac{\rm{Area}(\partial \mathcal{I})}{4 G_N^{(d+1)}} + S_{\rm{bulk}} (\mathcal{R} \cup \mathcal{I}), \label{Sgen}
\end{equation}
where $\mathcal{R}$ is the radiation region, $\mathcal{I}$ is the island, and $\mathcal{\partial I}$ is the boundary of the island which is a codimension-two surface and plays the role of the QES. The first term is a classical area term involving the boundary of the island, while the second term is the bulk entropy of the union of the radiation region and the island. The bulk entropy can be calculated in a scheme known as double holography. The fine-grained entropy of the Hawking radiation can be computed by extremizing the generalized entropy and if there are multiple extremal solutions, we choose the one that minimizes the generalized entropy,

\begin{equation}
  S_{\rm{R}} = \underset{I}{\rm{min}} \{ \underset{I}{\rm{ext}} \left[S_{\rm{gen}} \right]\}, \label{S fine-grained}
\end{equation}

At early times, there is no island contribution and so the first term of Eq.\eqref{Sgen} vanishes. This vanishing island contribution corresponds to the increase of the fine-grained entropy and eventually exceeds the coarse-grained entropy which then implies the information paradox. However, at a later time, a phase transition occurs with an island emerging and contributing to Eq.\eqref{Sgen}. The emergence of this island allows for the decrease of the fine-grained entropy of Hawking radiation through the extra dimensions. The increase and the eventual decrease of the fine-grained entropy are encapsulated by Eq.\eqref{S fine-grained}. The purification may be realized through the extra dimensions that is in the spirit of ER=EPR, and the entanglement wedge includes both the outgoing Hawking radiation and the black hole interior.

The island formulation was first utilized to study the Page curve in a 2d Jackiw-Teitelboim (JT) gravity that is coupled to a thermal bath \cite{1905.08762,1911.03402,2004.14944}. It was eventually extended to higher dimensional spacetime \cite{1910.12836,1911.09666,2004.05863,2005.08715,2006.02438,2006.04851,2006.10754,2006.11717,2006.16289,2007.06551,2007.15999,2010.00018,2010.00037,2010.04134,2010.16398,2012.03983,2012.05425,2012.07612,2101.05879,2101.10031,2101.12529,2102.01810,2102.02425,2103.15852,2103.16163,2104.00006,2104.02801,2104.07039,2104.00224,2104.13383,2105.00008,2105.01130,2105.08396,2105.09106,2106.07845,2106.10287,2106.11179,2106.12593,2107.00022,2107.01219,2107.03390,2111.14551,2206.09609,2211.13415,2304.09909,2306.12476}, with its validity studied extensively in \cite{2003.05448, 2003.11870, 2004.13857, 2004.14944, 2007.04877, 2007.10523, 2101.06867, 2103.14364, 2104.00052, 2105.12211, 2106.14738, 2107.03031, 2107.05218, 2107.07444, 2108.10144, 2109.07842, 2110.07598, 2112.06967, 2205.07905, 2305.04259}. The investigation of the QES in a higher dimensional thermofield double state using the Randall-Sandrum (RS) brane scenario has been examined in \cite{1911.09666,2006.02438}.
Eventually, the intrinsic brane curvature is introduced to the action of the modified braneworld theory \cite{2006.04851} to incorporate both the RS brane scenario and Dvali-Gabadadze-Porrati (DGP) gravity \cite{0005016}.

In our previous work \cite{2111.14551}, the island formulation is applied in the AdS-Schwarzschild spacetime under the context of boundary conformal field theory (BCFT) which is a conformal field theory with boundaries having appropriate boundary conditions imposed \cite{9505127}. In that setup, an eternal black hole coupled to a thermal bath was considered such that the Hawking radiation from the black hole could leak into the bath. The radiation front then forms a time-dependent hypersurface which we called a time-dependent end-of-the-world (ETW) brane that leads to a new phase of RT surface known as the \textit{BRT surface}, it is important to note that a more accurate name for the brane that is formed will be discussed below in order to distinguish it from the typical static ETW brane considered in most of the literature. The time-dependent finite region between the boundary and the brane contains the actual radiation from the black hole, and we call it the \textit{effective Hawking radiation region}. The calculations were done in a time-independent black hole spacetime at fixed temperatures.

For a real black hole evaporation, however, the black hole temperature essentially evolves through time and is therefore sensitive to the quantum backreaction. As shown in \cite{1606.03438,2304.04427} for evaporating $\rm{AdS}_2$ black holes, given some black hole temperature at early times, the temperature should eventually decay and vanish at the end of the evaporation. This implies that at the end of the evaporation, the temperature of the black hole vanishes, and assuming a quasi-equilibrium system the temperature of the radiation should also vanish. In this sense, it becomes important to consider the phase transition of the HRT surfaces that incorporates the corresponding time dependence of the temperature. On the other hand, a toy model that can be examined is where the time dependence is considered in a quasi-static scenario wherein the evaporation process is extremely slow.

%%%%%%%%%%%%%%%%
\textbf{\large Motivation and Discussion}

To date, most of the research addressing the information paradox for dimensions $d\geq3$ has predominantly concentrated on the study of eternal black holes, as in a series of important works \cite{2004.05863, 2010.00037, 2011.08814, 2101.06867, 2109.02996, 2110.04233, 2111.14551, 2112.14361, 2203.06310}. However, recent developments have focused on understanding the implications of Hawking radiation within time-dependent backgrounds. The work presented in \cite{2302.02379} explored the Page curve in a step-function Vaidya spacetime, offering a concrete framework to model black hole evaporation. Concurrently, the investigation presented in \cite{2301.02587} introduced elements of dynamical evolution on the brane, albeit within a static spacetime construct.

Our study contributes to this evolving topic by examining the Page curve through the lens of a fully time-dependent spacetime geometry, diverging from the conventional focus on static spacetimes, the partial dynamical evolutions \cite{2301.02587}, or the quench scenarios \cite{2302.02379}. We specifically investigate an evaporating black hole within the AdS-Vaidya spacetime, the characteristics of which are showed in the Penrose diagram Fig.\ref{Penrose-Vaidya}. This model illustrates the gradual evaporation of an AdS black hole, theoretically extending into infinity, as per the findings in \cite{1403.4886}. The dynamical mass of the evaporating black hole within the AdS-Vaidya spacetime framework is meticulously derived from the Stefan-Boltzmann law, emphasizing a time-dependent mass parameter.

On the other hand, in exploring the Page curve for evaporating black holes, the gravitational region of the system is purified through its interaction with a CFT bath. This bath, representing the Hawking radiation region, is conventionally considered infinite and capable of carrying limitless information. Yet, an alternative perspective, potentially more aligned with realistic scenarios, considers a finite-sized bath endowed with a limited spectrum of degrees of freedom.

The investigation into the Page curve within the context of a static spacetime, incorporating a bath of finite extent, has previously been studied in the literature \cite{2007.11658, 2202.00679,2209.15477}, where the definitive parameters indicative of the bath's dimensions remain constant. The emergence of an island phase within these models depends on the dimensional parameters of the finite baths. Within such frameworks, the dimensional parameter of the bath has to be manually adjusted to facilitate the manifestation of an island, thereby enabling the reconstruction of the black hole's interior. 

In contrast to preceding models that exclusively considered baths of fixed dimensions, we introduce a novel paradigm: a dynamically expanding CFT bath, termed the \textit{effective Hawking radiation region}. This concept utilizes time as a defining parameter for the bath's dimensions, representing a significant advancement over previous models.
To elucidate the concept of an effective Hawking radiation region, we have derived a non-trivial solution for a dynamic boundary condition, which we have designated as the \textit{moving-end-of-the-radiation} (METR) brane, in the framework of BCFT. This METR brane dynamically recedes from the black hole at the speed of light, effectively representing the advancing frontier of Hawking radiation. Within the bulk framework, the METR brane is aligned approximately along the light cone, in contrast to the static ETW brane, which is oriented parallel to the temporal axis, as illustrated in Fig.\ref{Different Branes}. It is necessary to highlight the temporal dependence of this new BCFT solution, which marks a significant departure from the predominantly static ETW brane configurations extensively discussed in prior literature.

Furthermore, the introduction of the METR brane facilitates the definition of a new type of HRT surface, which we refer to as the Boundary HRT (BHRT) surface. This BHRT surface plays a crucial role in shaping the structure of the Page curve during the initial evaporation phases. It is important to emphasize the fundamental distinction between the BHRT surface, as introduced in this study, and the extremal surfaces associated with a fixed finite bath as studied in the preceding research \cite{2007.11658, 2202.00679,2209.15477}. Their functional roles and behaviors are distinctively different.

In summary, our study carefully examines the fine-grained entropy of Hawking radiation through the island formulation within a BCFT framework, applying it to the AdS-Vaidya model of evaporating black holes. We have derived a novel non-trivial BCFT solution that captures the time-dependent dynamics of both spacetime and the surrounding bath's evolution. Our analysis focuses on the dynamic interplay between these elements, particularly the phase structure shaped by the competing dynamics of bath expansion and black hole evaporation. The introduction of the METR brane and the novel BHRT surface that dominates the early stages, compared to the IHRT surface associated with the Planck brane that dominates the late stages, delineates a consistent trajectory in the evolution of entanglement entropy. Notably, the Page time is shown to be dependent on the ratio of the bath's expansion rate to the black hole evaporation rate, which is itself proportional to the black hole mass.

Our approach advances the understanding of the intricate dynamics of black hole evaporation and provides a robust foundation in both classical and quantum gravitational theories.

The paper is organized as follows: In section \ref{Evaporating Black Hole}, we elaborate on the setup for evaporating black holes including its Penrose diagram. Section \ref{Boundary Conformal Field Theory} explains the BCFT setup for the evaporating black hole. Here, we emphasize on the significance of the METR brane, contrasting it with the static ETW brane that has been extensively studied in the literature. Additionally, we compare the METR brane discussed herein with that from our prior work \cite{2111.14551}. In section \ref{Holographic Entanglement Entropy}, we present the calculation of holographic entanglement entropies and the numerical methodologies employed. Section \ref{Page Curve} presents our findings, including the Page curve and the phase diagram. We culminate with a conclusion of our findings in section \ref{Conclusion}.

%%%%%%%%%%%%%%%%%%%%%%%%%%%%%%%%
\section{Evaporating Black Hole} \label{Evaporating Black Hole} 

To date, most works on the information problem for dimensions $d\geq3$ have been focused on eternal black holes \cite{2004.05863, 2010.00037, 2011.08814, 2101.06867, 2109.02996, 2110.04233, 2111.14551, 2112.14361, 2203.06310}. In addition, even though the model in \cite{2301.02587} has considered some element of dynamical evolution on the brane, the bulk is still essentially a static spacetime. On the other hand, in \cite{2302.02379}, the Page curve was studied under the step-function Vaidya spacetime to model the evaporation of a black hole. An s-wave approximation was used to study a one-sided black hole that is formed from the collapse of a positive mass null shell and the black hole vanishes through the collapse of a negative mass null shell. At late times, a Minkowski region is glued which serves as the scenario when the black hole completely vanishes. As they have discussed, their setup can partially model the evaporation of a black hole.

In our previous work \cite{2111.14551}, we studied a time-independent AdS-Schwarzschild black hole where the island formulation was considered under the context of BCFT. This is reviewed briefly in Appendix \ref{Appendix Eternal Black Hole}.

In the current work, a fully time-dependent spacetime is studied in order to investigate an evaporating black hole.  We consider an eternal AdS-Schwarzschild black hole that existed for an extremely long time in the past. At a certain time, the eternal black hole is allowed to evaporate which proceeds as ingoing negative energy flux allows for the black hole area to decrease, while the corresponding outgoing Hawking partners escape to infinity. The later time-dependent spacetime is then described by an outgoing AdS-Vaidya black hole. On the other hand, to examine the energy condition, we assume that the eternal AdS-Schwarzschild black hole was formed from a pure AdS spacetime by injecting a collapsing null shell. Some studies concerning black hole collapse in the context of AdS/CFT are \cite{0112099,0605224,0610041}.

\subsection{AdS-Vaidya Spacetime}
\begin{figure}
  \centering
  \includegraphics[scale=0.8]{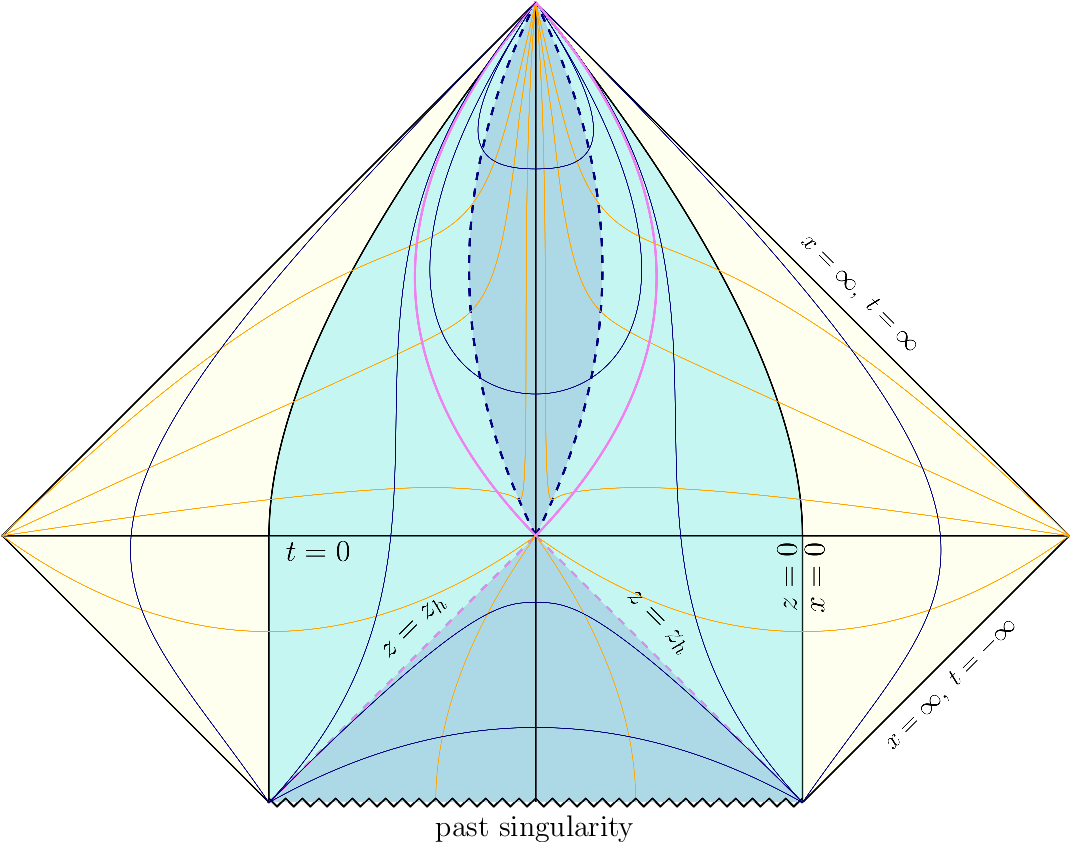}
  \caption{Penrose diagram of an evaporating black hole. The orange lines are the constant time slices, the navy blue solid lines are the constant space slices, the navy blue dashed lines are the apparent horizon, and the violet lines (solid and dashed) are locations where $z=z_{h}$.}
  \label{Penrose-Vaidya}
\end{figure}

In order to investigate the Page curve of an evaporating black hole characterized by a time-dependent spacetime, we analyze the $(d+2)$-dimensional outgoing AdS-Vaidya metric to represent the bulk spacetime $\mathcal{M}$,
\begin{equation} \label{Outgoing AdS Vaidya Metric}
  ds^2_{\mathcal{M}} = \frac{l^2_{AdS}}{z^2} \left[ -f(u,z) du^2 + 2dudz + dx^2 + \sum_{i=1}^{d-1} (dx_i)^2  \right],
\end{equation}
where $l_{AdS}$ is the AdS radius and
\begin{equation} \label{blackening factor}
  f(u,z) = 1 - m(u) z^{d+1}.
\end{equation}
The outgoing light-cone coordinate $u$ is defined as
\begin{equation}\label{Outgoing coordinate transformation}
  u = t + \int \frac{dz}{f(u,z)}.
\end{equation}
In this study, we focus on a double-sided black hole spacetime whose Penrose diagram illustrating the time-dependent spacetime under consideration is displayed in Fig.\ref{Penrose-Vaidya}. The regions illustrated in dark blue correspond to the spacetime that exists beyond the event or apparent horizon, while the light blue regions denote the asymptotic AdS spacetimes, with their boundaries located at $z=0$. The union of these dark and light blue areas is referred to as the gravitational region.

To explore black hole evaporation, we incorporate flat thermal baths onto the gravitational region by gluing them together at the coordinates $z=0$ and $x=0$ where $t=t_h$ is the time the first Hawking radiation allowed to leak into the bath reaches the boundary at $z = 0$. Herein we set $t_h = 0$ without the loss of generality. For time $t<0$, the boundary conditions at these coordinates impose a reflection condition, leading to the decoupling of the gravitational region and the thermal baths. However, for $t>0$, we apply the transparent condition at the boundary which initiates the evaporation of the black hole.

For $t<0$, the spacetime is characterized by an AdS-Schwarzschild black hole which is on the lower portion of the Penrose diagram, with the reflection boundary condition at $z=0$. At $t=0$, the black hole starts to evaporate and is described by the outgoing AdS-Vaidya spacetime given by Eq.\eqref{Outgoing AdS Vaidya Metric} which is on the upper portion of the Penrose diagram where the boundary condition transitions to a transparent condition, allowing the radiation from the black hole to enter the thermal bath region. In addition, the boundary $z=0$ separating the gravitational region and the thermal bath region becomes a curve in the Penrose diagram for $t>0$. In an effort to maintain control over the process, we postulate that the black hole undergoes an extremely slow evaporation. This assumption ensures that the gravitational region is consistently in equilibrium with the thermal baths throughout the entirety of the evaporation process. During evaporation, the size of the black hole decreases and the apparent horizon appears as a time-like curve depicted by the navy blue dashed line in Fig.\ref{Penrose-Vaidya}.

In Fig.\ref{Penrose-Vaidya}, the orange curves represent constant $t$ slices, while the navy blue curves correspond to constant  $r$ (here $r$ may correspond to $z$ or $x$ depending on the specific spacetime region) slices. The event horizon at $z=z_h$ for $t<0$ is delineated by the violet dashed lines. For $t>0$, the violet solid curves signify constant $z=z_h$ slices that exist outside of the apparent horizon which is represented by the navy blue dashed trajectories. For $t<0$, these slices exhibit the same behavior as in a static black hole. However, for $t>0$, the dynamics of the spacetime introduce modifications to these slices. In any region of spacetime, the constant $t$ curves behave as spacelike outside the horizon and timelike inside the horizon. On the other hand, the constant $r$ curves display timelike properties outside the horizon and spacelike features inside the horizon. Nevertheless, due to the presence of an event or apparent horizon, the constant $r$ slices for $z>z_h$ are distinctly separated between $t>0$ and $t<0$. These distinctions ensure the preservation of causality throughout the entire spacetime.

According to the mass function in Eq.(\ref{mass}) that describes the time-dependent AdS-Vaidya spacetime studied in this work (see also \cite{1403.4886,1503.08245}), the black hole will be completely evaporated as time tends towards infinity where as the mass of the black hole approaches zero there is no singularity present for $t>0$. Alternatively, one may view this as the singularity converging to a point at $t=\infty$ and $z=\infty$.

Given that the black hole undergoes complete evaporation as time approaches infinity, any matter or information contained within the black hole will inevitably escape at a finite time, as depicted in Fig.\ref{Penrose-Vaidya}. Notably, each photon entering the black hole from one side will eventually traverse to the opposite side within a finite time. This phenomenon enables intercommunication between the two sides in this configuration.

\subsection{Null Energy Condition (NEC)}
It is known that the local energy conditions in general relativity are allowed to be violated in quantum field theory, for which those conditions are utilized for the proof of singularity theorems, topological censorship theorem, etc. One of the conditions that could be violated is the null energy condition (NEC) which states that,
\begin{equation}
    T_{\mu\nu} k^\mu k^\nu \geq 0,
\end{equation}
for all null vectors $k^\mu$.

An improved condition that is formulated non-locally is the averaged null energy condition (ANEC),
\begin{equation}
    \int^\infty_{-\infty} \langle T_{\mu\nu} \rangle k^\mu k^\nu d\lambda \geq 0,
\end{equation}
where the integral is examined over an affinely parameterized null geodesic with $\lambda$ being the affine parameter. However, it has been demonstrated that ANEC can also be violated for any chronal null geodesic in Schwarzschild spacetime studied for a conformally coupled scalar field \cite{9604008}. Due to this, the achronal ANEC was proposed which supposes that ANEC holds for all complete achronal null geodesics \cite{0705.3193}.

In the context of AdS/CFT, it has been shown that the achronal ANEC can be violated for strongly coupled theories in curved spacetime \cite{1903.11806}. It was argued in \cite{1911.02654} that since ANEC is not conformally invariant, conformal transformations can magnify the local NEC violation at a given point. This motivated the proposal for a conformally invariant averaged null energy condition (CANEC) which holds on an achronal null geodesic segment for a class of strongly coupled field theories. Specifically, utilizing Gao-Wald's \textit{no-bulk-shortcut} property \cite{0007021}, the ANEC with a corresponding weight factor examined for spatially compact universes was shown to hold along null geodesics while upholding conformal invariance. In \cite{2111.05151}, a holographic model is considered whose boundary is an asymptotically flat evaporating black hole spacetime. The no-bulk-shortcut property was applied and it was shown that the ANEC together with a weight factor holds when the null energy is averaged along the incomplete null geodesic on the event horizon.

In this paper, a similar strategy is applied to show that ANEC along with a corresponding weight factor is satisfied for the evaporating background considered. For the purposes of examining ANEC along the null geodesic on the horizon in which the ingoing negative energy flux allows for the area decrease, it is suitable to use the ingoing AdS-Vaidya metric. On the contrary, the main aim of this paper is to examine the Page curve of Hawking radiation for which these modes are away from the horizon and the outgoing AdS-Vaidya metric is used. The ingoing AdS-Vaidya metric can be complemented by the outgoing AdS-Vaidya metric by employing a matching technique for the two metrics at the pair creation location. However, this is not needed for our purposes.

The strategy is to construct a 6-dimensional bulk spacetime dual to a 5-dimensional asymptotically AdS evaporating black hole. As stated in \cite{1911.02654}, the no-bulk-shortcut property implies that CANEC holds on the boundary so it is expected that a boundary field theory that violates CANEC cannot have a compatible holographic bulk dual. Considering the AdS-Vaidya metric as the boundary geometry, the bulk can at least be constructed locally near the boundary by doing a Fefferman-Graham expansion into the bulk. Recalling the no-bulk-shortcut property, suppose that an achronal null geodesic $\alpha$ on the boundary connects two boundary points, then the no-bulk-shortcut property implies that there exists no bulk timelike curve that connects those two boundary points which signifies that $\alpha$ is the fastest causal curve connecting the boundary points. Now, given a boundary achronal null geodesic $\alpha$, consider all bulk causal curves $\beta$ connected with $\alpha$ via Jacobi fields. In this sense, a neighborhood of curves is formed with respect to $\alpha$ covered by the Fefferman-Graham coordinates. Employing the no-bulk-shortcut property leads to the notion that $\alpha$ is the fastest causal curve in the neighborhood. The no-bulk-shortcut property will be used to obtain an inequality which will then lead to ANEC being shown to be satisfied with a corresponding weight factor. The calculations of these will be relegated to the Appendix \ref{Appendix ANEC}.

\subsection{Thermal Bath}
In investigating the Page curve for evaporating black holes, the black hole system is typically coupled to a CFT bath of infinite size for which the bath has the capacity to carry an infinite constituent of information. An alternate and presumably realistic consideration is to examine a bath of finite size that has finite degrees of freedom, e.g. see \cite{2202.00679,2007.11658,2209.15477}. 

In \cite{2202.00679}, the Page curve for a static spacetime with finite bath was studied where the endpoint that characterizes the bath size is fixed. In general, depending on how the finite bath size is tuned, there could be different extremal surfaces that may dominate in the course of the evolution. On the other hand, at least in the setup of \cite{2202.00679}, it was shown that there are two extremal surfaces that may dominate at late times. First, there is an extremal surface that is attached to the endpoint of the CFT bath that characterizes its size, which goes around and hugs the horizon, this surface corresponds to the thermal entropy of the CFT bath and is reflected in the Page curve as a constant curve. Second, there is an extremal surface that corresponds to the island contribution where the surface attaches to the CFT bath on one end and to the Planck brane on the other end where the black hole is located, it is reflected in the Page curve as a constant curve. For small bath regions, it was eventually found that the late-time phase should go to the extremal surface that goes around the horizon which gives a saturation of the entanglement entropy and avoids quantum extremal island formation. For large enough baths, the late-time phase is associated with the extremal surface corresponding to the island contribution hence allowing black hole interior reconstruction. In this setup, the bath parameter characterizing its size must be tuned properly in order for the island to emerge and therefore resolve the information paradox.

There were also studies about finite baths involving shock waves due to quench scenario \cite{2007.11658, 2209.15477}. The finite bath was again fixed by endpoints that characterize its size. Depending on the size of the bath and its relative position to the shock wave point, there is either no way for black hole interior reconstruction, a small window of time for the interior reconstruction, or a stable interior reconstruction. Overall, the bath size must be tuned to allow for black hole interior reconstruction.

In \cite{2301.02587}, a black string in AdS was considered to model an evaporating black hole where the bulk black hole whose horizon intersects the brane and the bulk horizon slides off the brane thus allowing the brane horizon to shrink and therefore modeling brane black hole evaporation. In this setup, even though the brane horizon shrinks, the bulk evolution is classical so the bulk horizon does not shrink and the bulk black hole does not disappear into bulk radiation, so in a sense, it is still eternal since the evaporation only occurs in the brane point of view. Even so, their setup still introduces some time-dependence and so there is a dynamical evolution. Due to this, it is still possible to calculate a Page curve based on the HRT prescription, however, they have refrained from doing so since they claimed that the numerical calculations were delicate and outside their scope. On the other hand, they gave some qualitative discussions on how the Page curve should behave in their model. They have discussed several possible holographic evaporation including twin black hole evaporation, a black hole radiating into another, etc. In their setup, they have also discussed the implications of a finite bath. As in the previous studies \cite{2007.11658,2202.00679,2209.15477}, the extremal surface that is attached to the endpoint that characterizes the fixed bath size plays a minor role since they discussed that this varies in time only if the whole geometry is time-dependent and even in that scenario, the variation is expected to be small. They discussed that their setup should be tailored so that the bath is large enough in order to avoid the transition to this extremal surface since this prevents quantum extremal islands from forming.

In contrast to preceding models that exclusively considered baths of fixed dimensions, in this work, we introduce a novel paradigm: a dynamically expanding CFT bath, termed the effective
Hawking radiation region. The details will be explained in the following sections.

%%%%%%%%%%%%%%%%%%%%%%%%%%%%%%%%
\section{Boundary Conformal Field Theory} \label{Boundary Conformal Field Theory}

The generalized entanglement entropy of a radiation region can be computed utilizing the island formulation by considering the boundary conformal field theory (BCFT) and doubly holographic techniques. As demonstrated in \cite{2006.02438}, three equivalent perspectives can be employed to describe the system.

\begin{figure}[t]
  \centering
  \begin{subfigure}{0.3\textwidth}
    \centering
    \includegraphics[height=0.35\textheight]{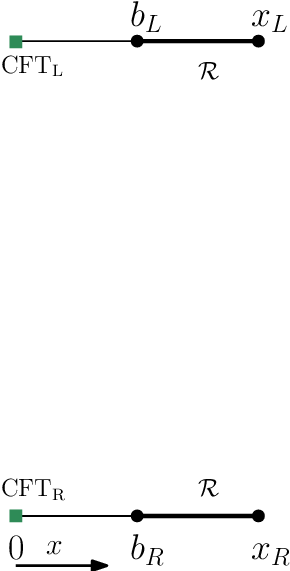}
    \caption{Boundary Perspective}
  \end{subfigure}
  \begin{subfigure}{0.3\textwidth}
    \centering
    \includegraphics[height=0.35\textheight]{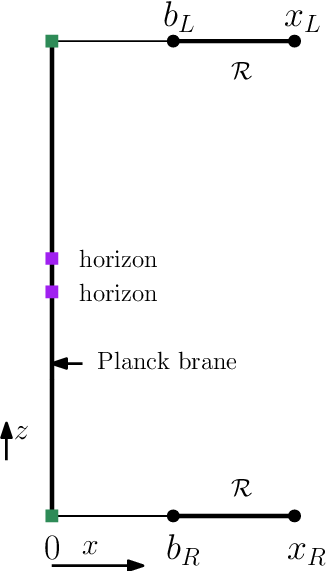}
    \caption{Brane Perspective}
  \end{subfigure}
  \begin{subfigure}{0.3\textwidth}
    \centering
    \includegraphics[height=0.35\textheight]{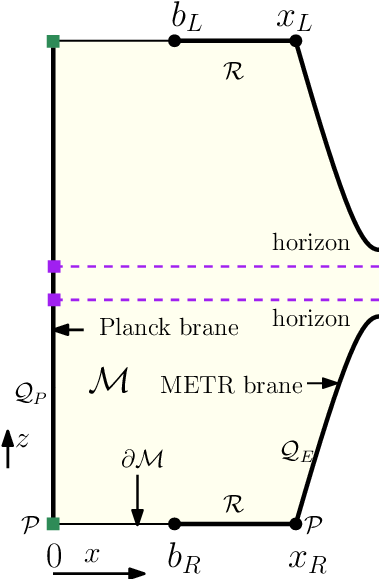}
    \caption{Bulk Perspective}
  \end{subfigure}
  \caption{Three perspectives of the BCFT setup. (a) A pair of $(d + 1)$-dimensional CFTs are situated on the region $[0, x_{L, R}]$ with two $d$-dimensional CFTs on their boundaries at $x=0$ indicated by green square dots, where $x = b_{L, R}$ are cutoffs and the other boundaries are at $x = x_{L, R}$. The effective Hawking radiation region $\mathcal{R}$ is indicated by $[b_L,x_L] \cup [b_R, x_R]$. (b) The $d$-dimensional CFTs are replaced by a Planck brane where an asymptotically $\rm AdS_{d+1}$-Vaidya black hole exists and the horizons are indicated by the purple square dots. (c) The $(d + 1)$-dimensional CFTs denoted by $\partial \mathcal{M}$ are replaced by an asymptotically $\rm AdS_{d+2}$-Vaidya black hole bulk spacetime $\mathcal{M}$ where the horizons are indicated by the purple dashed lines. The Planck brane is denoted by $\mathcal{Q}_P$ and the time-dependent METR brane is denoted by $\mathcal{Q}_E$. Due to the time-dependent AdS-Vaidya spacetime considered, the METR brane appears to be curved in the bulk spacetime.}
  \label{Three perspectives in BCFT}
\end{figure}

For the boundary perspective, a pair of $(d + 1)$-dimensional conformal field theories (CFTs) reside on a region $[0, x_{L, R}]$ with two $d$-dimensional CFTs living on their boundaries at $x = 0$ designated by the green square dots in Fig.\ref{Three perspectives in BCFT}(a). Here, $x = b_{L, R}$ represents the cutoff, and the other boundary is located at $x = x_{L, R}$. The effective Hawking radiation region is defined as $[b_L,x_L] \cup [b_R, x_R]$.

For the brane perspective, a Planck brane where an asymptotically $\rm AdS_{d+1}$-Vaidya black hole exists replaces the $\rm CFT_{d}$ by employing the holographic correspondence, as depicted in Fig.\ref{Three perspectives in BCFT}(b). We designate the radial direction of the black hole as $z$, with the $\rm CFT_{d}$ located at $z = 0$. The apparent horizons of the black hole are represented by two purple square dots.

For the bulk perspective, we once again use the holographic correspondence to replace the $\rm CFT_{d+1}$ with an asymptotically $\rm AdS_{d+2}$-Vaidya black hole bulk spacetime. The holographic BCFT setup accurately characterizes the geometry as presented in Fig.\ref{Three perspectives in BCFT}(c). This establishes the duality between the $\rm AdS_{d+2}$-Vaidya black hole bulk spacetime and the $(d + 1)$-dimensional BCFT. The Planck brane is established by protruding the boundary $x = 0$ into the bulk spacetime, while the METR brane is established by protruding the other boundary $x = x_{L, R}$ into the bulk spacetime. By solving the BCFT system, the embedding of the branes into the bulk spacetime can be determined. In the bulk black hole, the purple dashed lines represent the two apparent horizons. In this perspective, the classical HRT surface in the bulk spacetime anchored on the entangling surface at $x = b_{L, R}$ allows for the calculation of the bulk entanglement entropy in Eq. (\ref{Sgen}).
To summarize the setup, we consider a bulk spacetime $\mathcal{M}$ that is $(d+2)$-dimensional and is holographically dual to a CFT that is $(d+1)$-dimensional which is confined on the conformal boundary $\mathcal{\partial M}$, which in turn is bounded by $d$-dimensional boundaries $\mathcal{P}$. This system constitutes two $(d+1)$-dimensional hypersurfaces $\mathcal{Q}$ anchored at $\mathcal{P}$ where one is on the left at $x = 0$ corresponding to the Planck brane denoted by $\mathcal{Q}_P$, while the other one is on the right at $x = x_{L, R}$ corresponding to the METR brane denoted by $\mathcal{Q}_E$. The effective Hawking radiation region $\mathcal{R}$ is given by $[b_L,x_L] \cup [b_R, x_R]$ and the bulk is described by a time-dependent black hole. These are all shown in Fig.\ref{Three perspectives in BCFT}(c).

%%%%%%%%%%%%%%%%
\subsection{Branes in the BCFT System}

The total action of the bulk gravity theory is given by,
\begin{equation}
  S = S_{\mathcal{M}} + S_{GHY} + S_{\mathcal{Q}} + S_{\partial \mathcal{M}} + S_{\mathcal{P}},
\end{equation}
where
\begin{align}
  S_{\mathcal{M}} & = \int_{\mathcal{M}} \sqrt{-g} (R_{\mathcal{M}} - 2\Lambda_{\mathcal{M}}), \\
  S_{\mathcal{Q}} & = \int_{\mathcal{Q}} \sqrt{-h} (R_{\mathcal{Q}} - 2\Lambda_{\mathcal{Q}} + 2 K_{\mathcal{Q}}), \\
  S_{\partial \mathcal{M}} & = 2 \int_{\partial \mathcal{M}} \sqrt{-\gamma} K', \\
  S_{\mathcal{P}} & = 2 \int_{\mathcal{P}} \sqrt{-\sigma} \theta.
\end{align}
In the total action, $S_{\mathcal{M}}$ is the action of the bulk $\mathcal{M}$ where $R_{\mathcal{M}}$ is its intrinsic curvature and $\Lambda_{\mathcal{M}}$ is its cosmological constant, while the boundary term is described by the Gibbons-Hawking-York term $S_{GHY}$. The action of the $(d+1)$-dimensional hypersurface $\mathcal{Q}$ is described by $S_{\mathcal{Q}}$ where the induced metric is $h_{ab} = g_{ab} - n_{a}^{\mathcal{Q}} n_{b}^{\mathcal{Q}}$ and $n^{\mathcal{Q}}$ is its unit normal vector. The intrinsic curvature, the cosmological constant, and the trace of the extrinsic curvature of $\mathcal{Q}$ are given by $R_{\mathcal{Q}}$, $\Lambda_{\mathcal{Q}}$, $K_{\mathcal{Q}}$, respectively. The action of the conformal boundary $\partial \mathcal{M}$ is described by $S_{\partial \mathcal{M}}$ where the induced metric is $\gamma_{ab} = g_{ab} - n_{a}^{\partial \mathcal{M}} n_{b}^{\partial \mathcal{M}}$, $n^{\partial \mathcal{M}}$ is its unit normal vector, and $K'$ is the trace of its extrinsic curvature. The boundary term $\mathcal{P}$ of $\mathcal{Q}$ and $\partial \mathcal{M}$ is described by the action $S_{\mathcal{P}}$ where its metric is $\sigma_{ab}$, and $\theta = \cos^{-1} \left( n^{\mathcal{Q}} \cdot n^{\partial \mathcal{M}} \right)$ is the supplementary angle between $\mathcal{Q}$ and $\partial \mathcal{M}$ which makes a well-defined variational principle on $\mathcal{P}$.

The semi-classical approach is based on the fixed background spacetime approximation that is not in accord with energy conservation which necessitates further correction to the background spacetime. However, for black holes with not too high temperatures, it is a feasible assumption that the evaporation process is quasi-static such that the radiation of the black hole has a temperature described by,
\begin{equation}
  T = \frac{(d+1)}{4\pi} \cdot m^{1/d+1}. \label{AdS-Vaidya temperature}
\end{equation}
According to the Stefan-Boltzmann law, the mass loss is given by,
\begin{equation}
  \frac{dm}{du} = - \sigma T^{d+1}, \label{Stefan-Boltzmann law}
\end{equation}
where $\sigma$ is a proportionality constant \cite{1403.4886}. After substituting Eq.\eqref{AdS-Vaidya temperature} into Eq.\eqref{Stefan-Boltzmann law}, we obtain,
\begin{equation}
  \frac{dm}{du} = - \sigma m, \label{Mass loss}
\end{equation}
where we have  combined all the constants into $\sigma$. Solving the above equation results in the exponential decay mass function,
\begin{equation}
  m(u) = m_0 e^{- \sigma u}, \quad 0 \leq u < \infty \label{mass}
\end{equation}
where $m_0$ denotes the initial mass at $u=0$.

Going back to the action $S_{\mathcal{Q}}$ of the hypersurface $\mathcal{Q}$, varying this with respect to $h^{ab}$ we obtain the equations of motion given by,
\begin{equation}
  R_{\mathcal{Q} ab} + 2 K_{ab} - \left( \frac{1}{2} R_{\mathcal{Q}} + K_{\mathcal{Q}} - \Lambda_{\mathcal{Q}} \right) h_{ab} = 0,\label{Ncondition}
\end{equation}
which is the Neumann boundary condition proposed by Takayanagi in \cite{1105.5165}. However, due to this condition being too strong, i.e., it has more constraints than the degrees of freedom, a mixed boundary condition is proposed in \cite{1701.04275, 1701.07202} given by,
\begin{equation} \label{mixed BC}
  (d-1) (R_{\mathcal{Q}} + 2K_{\mathcal{Q}}) - 2(d+1) \Lambda_{\mathcal{Q}} = 0.
\end{equation}
As discussed, the BCFT system comprises two hypersurfaces $\mathcal{Q}_P$ and $\mathcal{Q}_E$ shown in Fig.\ref{Three perspectives in BCFT}(c). The Planck brane $\mathcal{Q}_P$ is time-independent and is described by the equation $x=0$. On the other hand, the METR brane $\mathcal{Q}_E$ is time-dependent and is described by the equation $x=u$. A time-dependent black hole solution of the BCFT system is given by the AdS-Vaidya black hole in Eq.\eqref{Outgoing AdS Vaidya Metric}.

The induced metric of $\mathcal{Q}_P$ and $\mathcal{Q}_E$ are,
\begin{eqnarray}
ds_{\mathcal{Q}_P}^2 &=& \frac{l^2_{AdS}}{z^2} \left[ -f(u,z)du^2 + 2dudz + \sum_{i=1}^{d-1} (dx_i)^2 \right], \\
 \nonumber \\
ds_{\mathcal{Q}_E}^2 &=& \frac{l^2_{AdS}}{z^2} \left[ \left( 1 - f(u,z) \right) du^2 + 2dudz + \sum_{i=1}^{d-1} (dx_i)^2 \right].
  \label{metric of ETW}
\end{eqnarray}
The intrinsic curvature, the trace of the extrinsic curvature, and the cosmological constant for the two hypersurfaces that satisfy the mixed boundary condition Eq.(\ref{mixed BC}) are given by,
\begin{eqnarray}
  R_{\mathcal{Q}_P} &=& -\frac{d(d+1)}{l_{AdS}^2}, ~ K_{\mathcal{Q}_P} = 0, ~ \Lambda_{\mathcal{Q}_P} = -\frac{d(d-1)}{2 l_{AdS}^2},\label{Planck brane solution}\\
 \nonumber \\
  R_{\mathcal{Q}_E} &=& 0, ~ K_{\mathcal{Q}_E} = -\frac{d+1}{l^2_{AdS}}, ~ \Lambda_{\mathcal{Q}_E} = -\frac{d-1}{l^2_{AdS}}.\label{ETW brane solution}
\end{eqnarray}

%%%%%%%%%%%%%%%%
\subsection{Moving End-of-The-Radiation (METR) Brane}

\begin{figure}[t]
\captionsetup[sub]{justification=centering}
  \begin{subfigure}[b]{0.32\textwidth}
    \includegraphics[width=\textwidth]{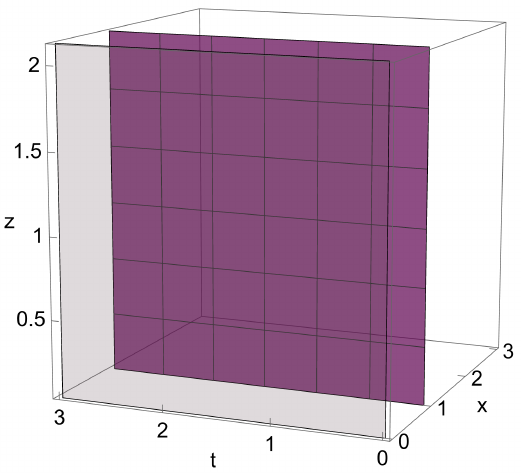}
    \caption{}
  \end{subfigure}
  \hfill
  \begin{subfigure}[b]{0.32\textwidth}
    \includegraphics[width=\textwidth]{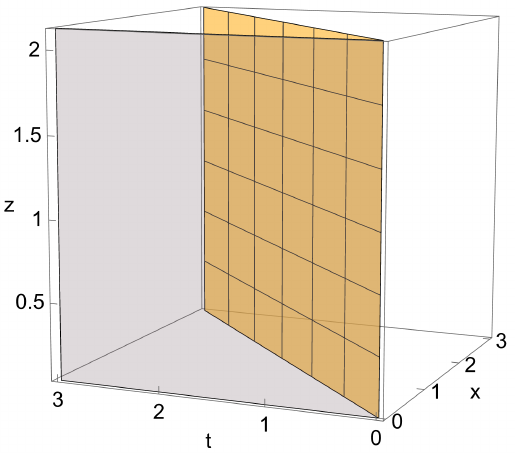}
    \caption{}
  \end{subfigure}
    \hfill
  \begin{subfigure}[b]{0.32\textwidth}
    \includegraphics[width=\textwidth]{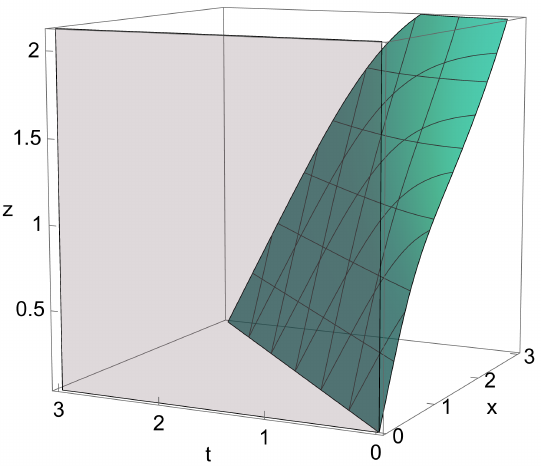}
    \caption{}
  \end{subfigure}
  \caption{Branes in different scenarios where the translucent gray colored plane represents the Planck brane located at $x=0$. (a) The static ETW brane located at $x=1$ is parallel to the $t$-axis. (b) The METR brane $x=t$ in the AdS-Schwarzschild spacetime studied in our previous work \cite{2111.14551} is tilted away from the $t$-axis. (c) The METR brane $x=u$ in the AdS-Vaidya spacetime is also tilted away from the $t$-axis, but in addition, it is tilted away from the $t-z$ plane and tilted towards the $x-z$ plane. This is clearly different from the static ETW brane and the METR brane in AdS-Schwarzschild spacetime.}
  \label{Different Branes}
\end{figure}

The conventional static ETW brane mostly considered in the preceding literature is illustrated in Fig.\ref{Different Branes}(a), where the typical RT surface in contact with the static ETW brane lies on a constant time slice. In our previous work \cite{2111.14551}, we explored an AdS-Schwarzschild spacetime that is time-independent. However, we derived a time-dependent BCFT solution, $x=ct$ (herein we set the speed of light to $c=1$). This led to a time-dependent METR brane, that is tilted away from the $t$-axis as depicted in Fig.\ref{Different Branes}(b). Nevertheless, the RT surface still lies on a constant time slice due to the temporal invariance of the bulk spacetime.
 
In the present work, we extend our focus to a time-dependent AdS-Vaidya spacetime. We derived a solution, illustrated in Eq.(\ref{metric of ETW}), for the METR brane within the BCFT framework. As depicted in Fig.\ref{Different Branes}(c), the new BCFT solution $x=u$ results in a time-dependent METR brane in the AdS-Vaidya spacetime which similarly tilts away from the $t$-axis. Moreover, the distinction of this configuration is that the new METR brane is not only tilted away from the $t$-axis, but also bends away from the $t-z$ plane. The quantities described in Eq.(\ref{ETW brane solution}) suggest that while the METR brane in AdS-Vaidya spacetime is intrinsically flat, its apparent curvature arises from its embedding within the bulk spacetime.  

In the vicinity of the conformal boundary $\partial \mathcal{M}$ at $z=0$, we impose a boundary condition on the METR brane that aligns to the light cone defined by $x=t$. This particular choice is based on the hypothesis that the Hawking radiation, which originates from the point where it enters the bath at $x=0$, propagates at light speed towards the asymptotic boundary at $x=\infty$.

The light cone demarcates the expanding region influenced by the Hawking radiation within the surrounding environment and defines the effective Hawking radiation region in the bath. This demarcation is graphically represented by an arrowed line in the Penrose diagram, illustrated in Fig.\ref{Penrose Diagram AdS-Vaidya HRT}.   On Cauchy surfaces at various times, the effective Hawking radiation regions are shown in red and blue segments, as presented in Fig.\ref{Penrose Diagram AdS-Vaidya HRT}.  The red segment represents the predominance of the BHRT surface associated with the METR brane. Over time, it undergoes a transition to the blue segment which symbolizes the emerging dominance of the IHRT surface linked to the Planck brane.

Our focus on the effective Hawking radiation region provides a crucial distinction from previous studies related to the Page curve. The introduction of the METR brane within the context of AdS-Vaidya spacetime offers a new perspective on the HRT surface, specifically the BHRT surface. This topic will be examined in detail in section \ref{Holographic Entanglement Entropy} of our study.

%%%%%%%%%%%%%%%%%%%%%%%%%%%%%%%%
\section{Holographic Entanglement Entropy} \label{Holographic Entanglement Entropy}

\begin{figure}[t]
  \begin{center}
    \includegraphics[width=0.9\textwidth]{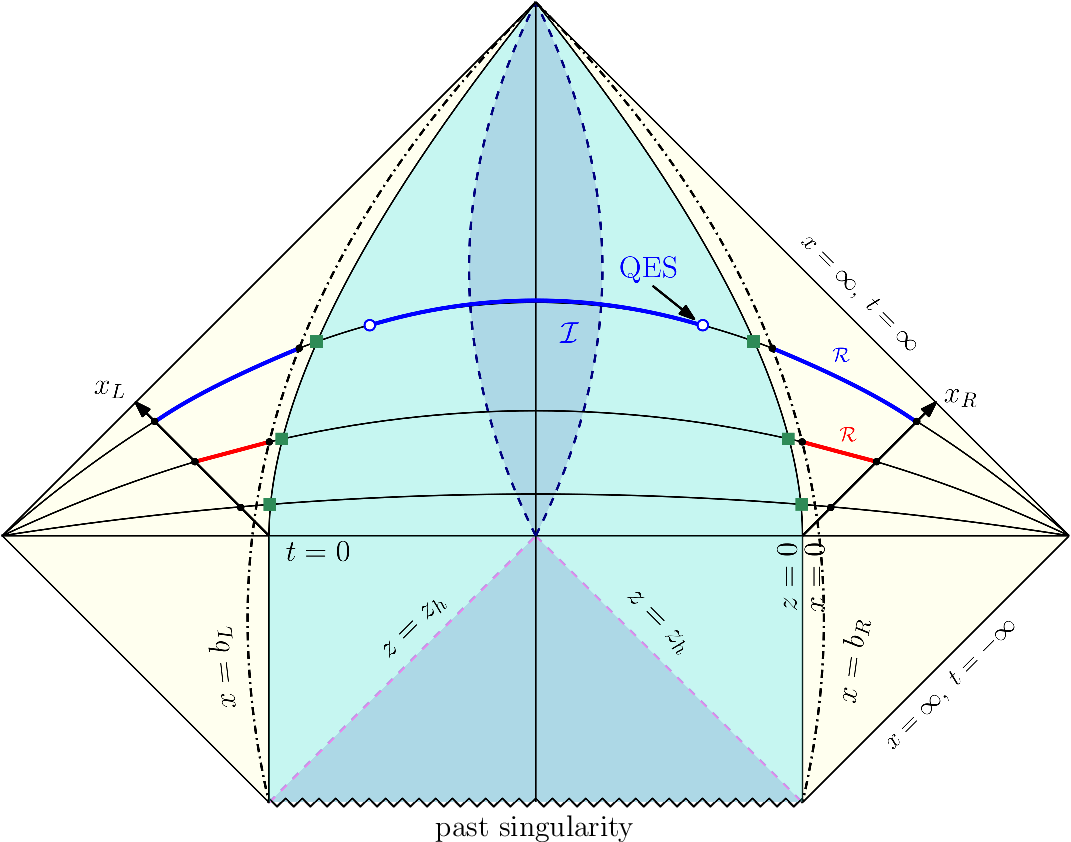}
    \caption{Penrose diagram of an evaporating black hole. As the radiation reaches the boundary $z=0$ at $t=0$, it leaks into the bath indicated by the yellow region, it eventually propagates toward the cutoff $x=b_R$ at $t=t_b$ indicated by the black dot-dashed lines, we then indicate $t_b$ as the initial time. After $t=t_b$, the effective Hawking radiation region grows in time and is dominated by the BHRT surface indicated by the red region. The effective Hawking radiation region continues to grow until a certain point where a phase transition occurs and is now dominated by the IHRT surface indicated by the blue region. The emergence of the island $\mathcal{I}$ along with the QES occurs during the IHRT surface domination.}
    \label{Penrose Diagram AdS-Vaidya HRT}
  \end{center}
\end{figure}

The choice of the entanglement region for the entanglement entropy calculation will be the effective Hawking radiation region $[b_L,x_L] \cup [b_R, x_R]$. Invoking the holographic correspondence in the doubly holographic bulk spacetime, the entanglement entropy is proportional to the area of the HRT surface which is anchored at the entanglement surface $x=b_{L,R}$ and is homologous to the effective Hawking radiation region. We only consider the classical HRT surface at the leading order such that the area of the surface is given by the Nambu-Goto string action,
\begin{equation} \label{Nambu-Goto Action}
  S = \int d^{d+1} \xi \mathcal{L} =\int d^{d+1} \xi \sqrt{\det G_{ab}},
\end{equation}
where the induced metric is defined by
\begin{equation}
  G_{ab}=g_{\mu\nu} \partial_a X^\mu \partial_b X^\nu,
\end{equation}
and the bulk metric $g_{\mu\nu}$ is given in Eq.\eqref{Outgoing AdS Vaidya Metric}.

In general, there could be more than one extremal surface that satisfies the homology constraint, in that case we choose the one that is minimal. In the system that we consider in this work, there exist two extremal surfaces. One is the BHRT surface which is anchored on the cutoffs at $x=b_{L,R}$ and also on the METR brane $\mathcal{Q}_E$. The other one is the IHRT surface which is anchored on the cutoffs $x=b_{L,R}$ and on the Planck brane $\mathcal{Q}_P$, their intersection of which is where the QES at $z=z_{\rm QES}$ is located.

\begin{figure}[t]
  \begin{center}
    \includegraphics[width=0.8\textwidth]{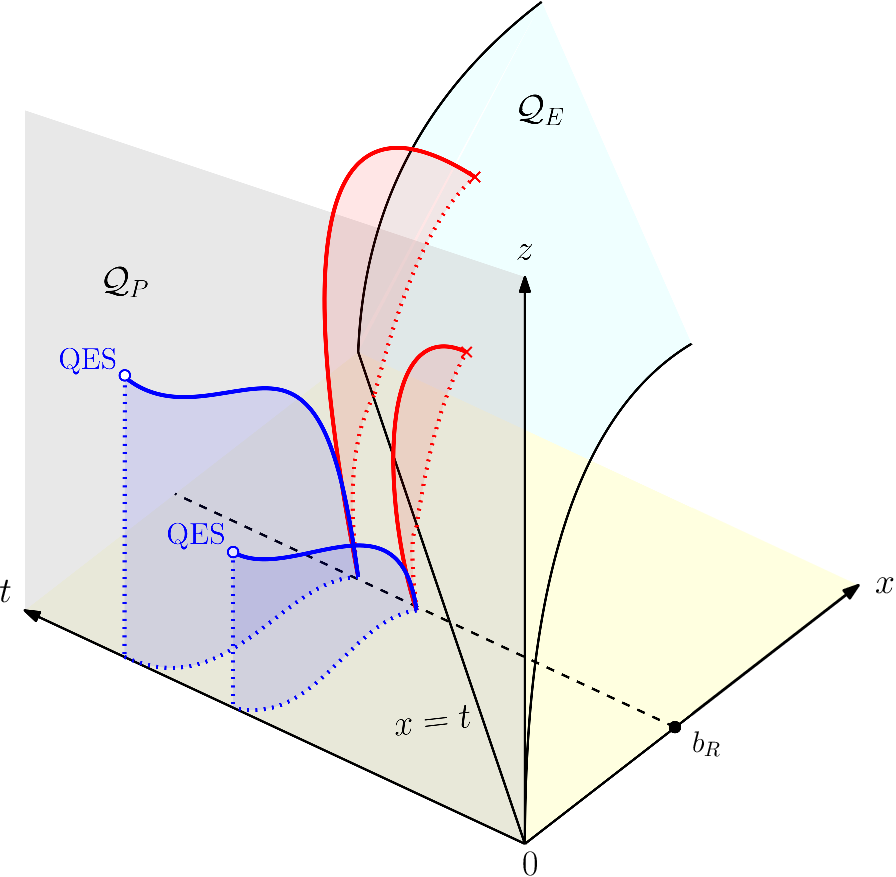}
    \caption{HRT surfaces in the bulk spacetime. The red curves represent BHRT surfaces, which satisfy the boundary condition on the METR brane $\mathcal{Q}_E$. The blue curves represent IHRT surfaces which then touch the Planck brane $\mathcal{Q}_P$ at the QES points and also satisfy the corresponding boundary conditions. It is important to note that the difference between the two HRT surfaces is that the IHRT surface is hosted on the time-independent Planck brane $\mathcal{Q}_P$, while the BHRT surface is hosted on the time-dependent METR brane $\mathcal{Q}_E$. The BHRT surface is nontrivial in the sense that on top of the time-dependent AdS-Vaidya spacetime considered, we also have a time-dependent radiation region defined by the METR brane.}
    \label{HRT Surfaces Bulk}
  \end{center}
\end{figure}

The emission of Hawking radiation from the black hole reaches the boundary at $z = 0$ at time $t=0$. This radiation subsequently propagates toward the cutoffs $x=b_{L,R}$ at time $t=t_b$, with $t_b$ signifying the beginning of the observation period. Following the passage of the Hawking radiation beyond the cutoffs $b_{L,R}$, the effective Hawking radiation region $[b_L,x_L] \cup [b_R, x_R]$ starts to expand. As the effective Hawking radiation region grows, the BHRT surface predominates the process until a critical time, i.e. Page time, is reached, at which point a phase transition occurs, leading to the dominance of the IHRT surface. The contribution of the BHRT surface at early times and the eventual emergence of the island $\mathcal{I}$ bounded by the QES at late times represented by the IHRT surface in the AdS-Vaidya spacetime is shown in the Penrose diagram in Fig.\ref{Penrose Diagram AdS-Vaidya HRT}. A sketch of these HRT surfaces at different times is shown in Fig.\ref{HRT Surfaces Bulk}. The red lines represent the BHRT surfaces that end on the METR brane satisfying the corresponding boundary conditions, while the blue lines represent IHRT surfaces that end on the Planck brane where the QES is located and satisfying the corresponding boundary conditions.

\begin{figure}[t]
\captionsetup[sub]{justification=centering}
  \begin{subfigure}[b]{0.49\textwidth}
    \centering
    \includegraphics[height=0.2\textheight]{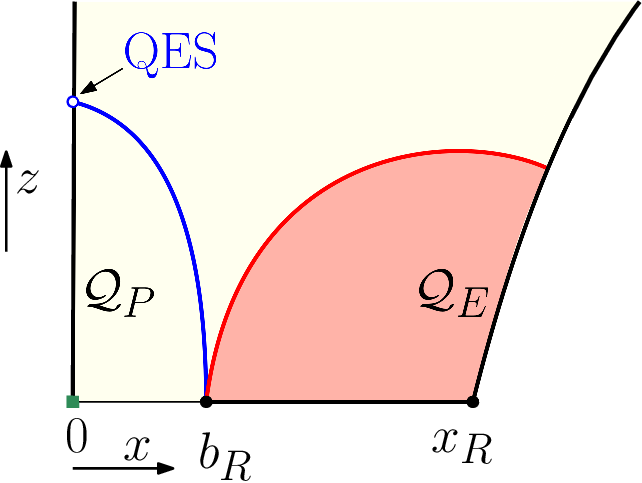}
    \caption{}
  \end{subfigure}
  \hfill
  \begin{subfigure}[b]{0.49\textwidth}
    \includegraphics[height=0.2\textheight]{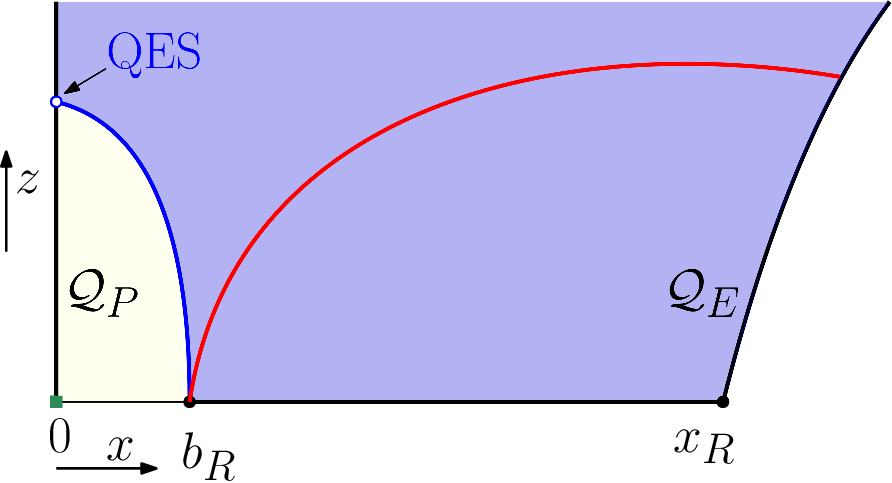}
    \caption{}
  \end{subfigure}
  \caption{Projected depictions of the HRT surfaces in the bulk spacetime where the BHRT surface (red curve) touches the METR brane $\mathcal{Q}_E$ and the IHRT surface (blue curve) touches the Planck brane $\mathcal{Q}_P$ at the QES. (a) Early time phase where the BHRT surface dominates with the entanglement wedge as the red-shaded region. (b) Late time phase where the dominance shifts to the IHRT surface with the entanglement wedge as the blue-shaded region.}
  \label{HRT Surfaces Projection}
\end{figure}

Fig.\ref{HRT Surfaces Projection} depicts the projection of the BHRT and IHRT surfaces onto the $z-x$ plane at certain times. At an earlier time, the BHRT surface dominates the system, with the red region representing the entanglement wedge as illustrated in Fig.\ref{HRT Surfaces Projection}(a). As time progresses, after the Page time, the BHRT surface surpasses the IHRT surface, causing the IHRT to become dominant. In this phase, the blue region denotes the entanglement wedge, which now includes the island region as displayed in Fig.\ref{HRT Surfaces Projection}(b). According to the principles of entanglement wedge reconstruction, the entirety of information contained within the black hole can be effectively reconstructed utilizing data observed in the effective Hawking radiation region after the Page time.

\subsection{Numerical Scheme}\label{Numerical Scheme}

For time-dependent spacetimes, we should use the HRT prescription. Typically, the nontrivial HRT surfaces can only be dealt with numerically. In our BCFT setup, the HRT surfaces are solved as geodesics in a boundary value problem which can be handled using relaxation methods \cite{Press2007}. In this method, differential equations are replaced by finite difference equations defined on a grid. An initial guess is prescribed to the system and the result relaxes to the true solution.

In the setup, we establish a grid $\sigma_i = h \, i$ of equal spacing $h = \frac{\sigma_n-\sigma_1}{n-1}$ where $i = 1, \ldots n$, and we use a grid that is composed of $n=500$ points. The HRT surfaces are parameterized in terms of the coordinates $(u, z, x)$. Given an approximate solution $r_{\rm{old}}$, we compute the increment $\Delta r$ such that $r_{\rm{new}} = r_{\rm{old}} + \omega \Delta r$ is an improvement to the old solution $r_{\rm{old}}$ where $\omega$ is the relaxation parameter. There is no obvious guideline in choosing the relaxation parameter, but faster convergence was achieved by choosing $\omega = 0.3$. The equations describing the HRT surfaces in the BCFT are nonlinear in nature and so the iteration can be done using Newton's method. It is useful to note that in addition to the equations being nonlinear, it is also coupled, and so the ordering of the equations when constructing the sparse matrix becomes crucial for faster convergence. The guess solution will typically not satisfy the governing equations as well as the boundary conditions accurately, and so in the calculations we can examine the discrepancy from the true solution using the residual error $\delta$. The iterations were done such that the residual error satisfies the tolerance $\delta < 10^{-10}$.

\subsection{BHRT Surface}\label{BHRT Surface}

The effective Hawking radiation region at a fixed time is,
\begin{equation}
  \begin{split}
    \mathcal{R} = \left\{ x \in (b_R, x_R), \textbf{x}_{d-1} \in \mathbb{R}^{d-1} \right\},
  \end{split}
\end{equation}
which preserves $(d-1)$-dimensional translation invariance. The BHRT surface can be described by $u=u(x)$ and $z=z(x)$, which leads to the following induced metric,
\begin{equation}
  ds^{2}=\frac{l_{AdS}^{2}}{z^{2}}\left[  \left( -f(u, z)
  u^{\prime2} + 2 u' z' + 1\right)  dx^{2}+\sum
  _{i=1}^{d-1}dx_{i}^{2}\right],
\end{equation}
where the prime denotes the derivative with respect to $x$.

In the following calculations, we are going to denote the $(d-1)$-dimensional volume as,
\begin{equation}
  \int d^{d-1} \mathbf{x} = V_{d-1}.
\end{equation}
The entanglement entropy becomes,
\begin{equation}
  \mathcal{S}_{\rm BHRT} = \frac{l_{AdS}^d V_{d-1}}{4 G_N^{(d+2)}} \int_{x_b}^{x_{\rm{R}}}dx  \frac{1}{z^d}
  \sqrt{-f(u, z) u^{\prime2} + 2 u' z' + 1}, \label{SBHRT-action}
\end{equation}
where $x_b$ is the CFT cutoff and $x_R$ is the $x$-coordinate of the radiation front. We extremize this to obtain the corresponding extremal surface.

For the time-dependent brane given by $u=x$, the boundary effect for the BHRT surface can be studied by reparameterizing it as $(u(\lambda), z(\lambda), x(\lambda))$. The action then reads,
\begin{equation}
  \mathcal{S}_{\rm BHRT} = \frac{l_{AdS}^d V_{d-1}}{4 G_N^{(d+2)}} \int_{\lambda_b}^{\lambda_R} d\lambda \frac{1}{z^d} \sqrt{-f(u,z) \dot{u}^2 + 2 \dot{u}\dot{z} + \dot{x}^2}, \label{SBHRT-action-affine}
\end{equation}
where the dot denotes the derivative with respect to the affine parameter $\lambda$. Varying the action Eq.\eqref{SBHRT-action-affine} we find that the boundary term is given by,
\begin{equation}
  \frac{\left( -f(u,z) \dot{u} + \dot{z} \right) \delta u + \dot{u} \delta z + \dot{x} \delta x}{z^d \sqrt{-f(u,z) \dot{u}^2 + 2 \dot{u}\dot{z} + \dot{x}^2}} \; \Biggr|_{\rm{boundary}}^{\textrm{METR brane}}.
\end{equation}
A Dirichlet condition is typically imposed on the entangling subregion, the calculation of which is done abundantly in the literature and will not be repeated here. The boundary condition of interest here is the one on the METR brane. For the variation to be well-defined, it must be the case that
\begin{equation}
  \left( -f(u,z) \dot{u} + \dot{z} \right) \delta u + \dot{u} \delta z + \dot{x} \delta x \Biggr|^{\textrm{METR brane}} = 0.
\end{equation}
\begin{figure}[t]
  \centering
  \includegraphics[scale=0.5]{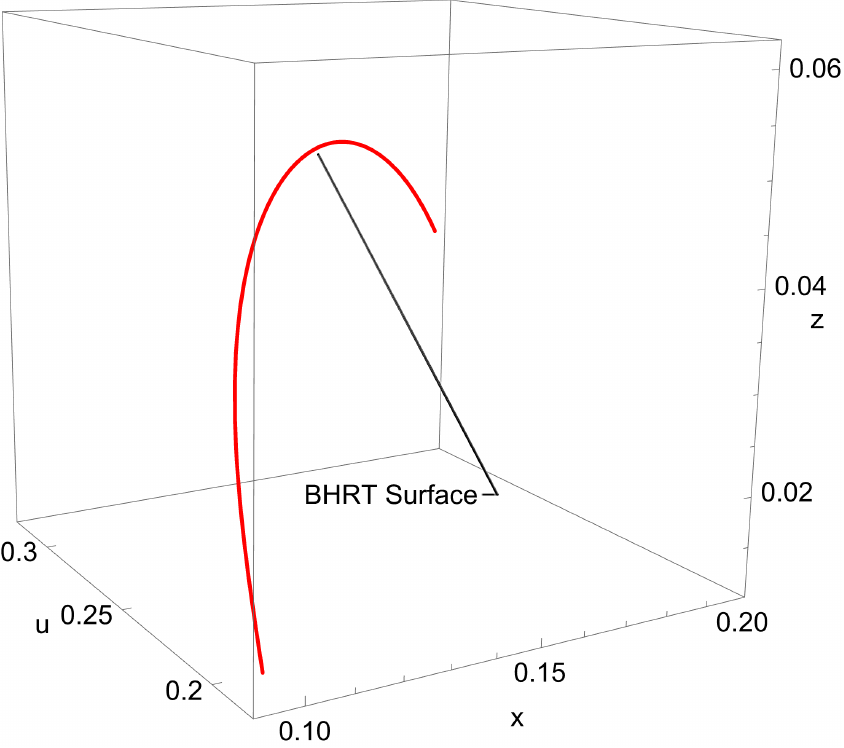}
  \caption{A sample solution for the BHRT surface in the AdS-Vaidya spacetime}
  \label{BHRT surface}
\end{figure}
The METR brane is given by the equation $x = u$ so that we require $\delta x = \delta u$. The above boundary condition then becomes,
\begin{equation}
  \left( -f(u,z) \dot{u} + \dot{z} + \dot{x} \right) \delta u + \dot{u} \delta z \Biggr|^{\textrm{METR brane}} = 0.
\end{equation}
The boundary condition in our case does not involve fixing a point on the METR brane, and so it is imperative that we have the equations,
\begin{flalign}
  \dot{u} & = 0\\
  -f(u,z) \dot{u} + \dot{z} + \dot{x} & = 0,
\end{flalign}
which then leads to,
\begin{flalign}
  & u' = \frac{du}{dx} = \frac{\dot{u}}{\dot{x}} = 0\\
  & z' = \frac{dz}{dx} = \frac{\dot{z}}{\dot{x}} = -1.
\end{flalign}
An example solution that shows how the BHRT surface behaves is shown in Fig.\ref{BHRT surface}. The figure clearly shows the nontrivial nature of the BHRT surface in contrast to the BRT surface \cite{2111.14551} which lies in the constant time slice.

\subsection{IHRT Surface}\label{IHRT Surface}

For the IHRT surface, the entanglement entropy is given by,
\begin{equation}
  \mathcal{S}_{\rm IHRT} = \frac{l_{AdS}^d V_{d-1}}{4 G_N^{(d+2)}} \int_{0}^{x_{\rm{b}}}dx  \frac{1}{z^d}
  \sqrt{-f(u, z) u^{\prime2} + 2 u' z' + 1} + \frac{1}{z_I^{d-1}},
  \label{SIHRT Action}
\end{equation}
where $x_b$ is the CFT cutoff and the QES contributes an additional term as compared to the BHRT case. The location of the QES on the Planck brane is denoted by $z_{\rm{QES}} = z_I$. The extremization can be done by rewriting the action Eq.\eqref{SIHRT Action} as,
\begin{equation}
  \mathcal{S}_{\rm IHRT} = \frac{l_{AdS}^d V_{d-1}}{4 G_N^{(d+2)}} \int_{0}^{x_{\rm{b}}}dx  \left[ \frac{1}{z^d} \left(
  \sqrt{-f(u, z) u^{\prime2} + 2 u' z' + 1} + (d-1) z' \right) \right] + \frac{1}{z_b^{d-1}}.
  \label{SIHRT Action Modified}
\end{equation}
where $z_b$ is the cutoff at the CFT boundary. To extract the boundary term, we rewrite the action as,
\begin{equation}
  \mathcal{S}_{\rm IHRT} = \frac{l_{AdS}^d V_{d-1}}{4 G_N^{(d+2)}} \int_{0}^{\lambda_b} d\lambda  \left[ \frac{1}{z^d} \left(
  \sqrt{-f(u,z) \dot{u}^2 + 2 \dot{u}\dot{z} + \dot{x}^2} + (d-1) \dot{z} \right) \right] + \frac{1}{z_b^{d-1}}.
  \label{SIHRT Action Affine}
\end{equation}
Varying the action Eq.\eqref{SIHRT Action Affine} with $1/z_b^{d-1}$ acting as an additional constant, we find that the boundary term is given by,
\begin{equation}
  \frac{\left( -f(u,z) \dot{u} + \dot{z} \right) \delta u +  \left( \dot{u} + (d-1) \sqrt{-f(u,z) \dot{u}^2 + 2 \dot{u}\dot{z} + \dot{x}^2} \right) \delta z + \dot{x} \delta x}{z^d \sqrt{-f(u,z) \dot{u}^2 + 2 \dot{u}\dot{z} + \dot{x}^2}} \; \Biggr|_{\textrm{Planck brane}}^{\rm{boundary}},
\end{equation}
where we require that,
\begin{equation}
  \left( -f(u,z) \dot{u} + \dot{z} \right) \delta u +  \left( \dot{u} + (d-1) \sqrt{-f(u,z) \dot{u}^2 + 2 \dot{u}\dot{z} + \dot{x}^2} \right) \delta z + \dot{x} \delta x \Biggr|_{\textrm{Planck brane}} = 0.
\end{equation}
\begin{figure}[t]
  \centering
  \includegraphics[scale=0.55]{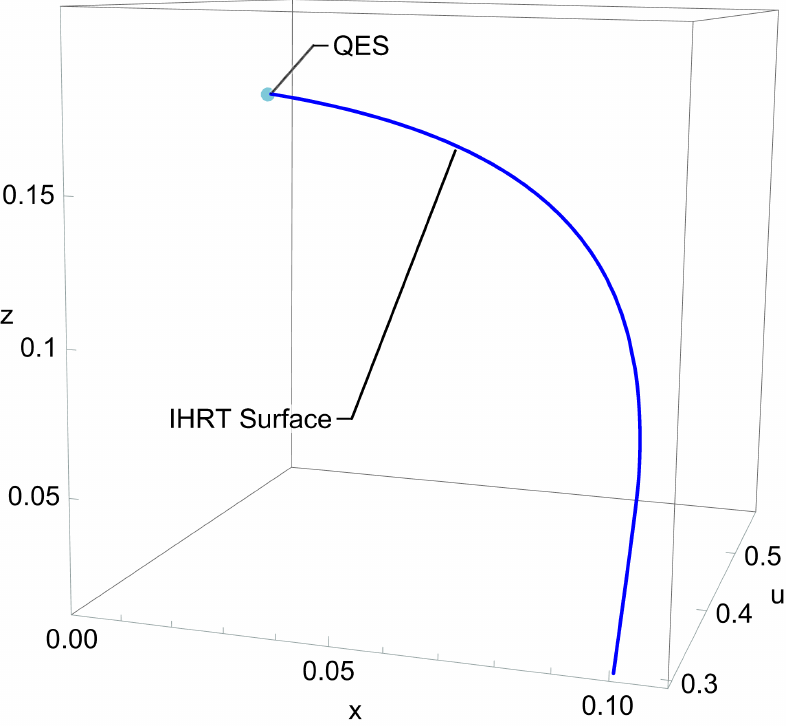}
  \caption{A sample solution for the IHRT surface in the AdS-Vaidya spacetime.}
  \label{IHRT surface}
\end{figure}
The boundary condition that we will consider in this case is the one on the Planck brane given by the equation $x=0$, and so we require that,
\begin{equation}
  \delta x = 0
\end{equation}
Again, allowing the boundary point to not be fixed leads to the equations,
\begin{flalign}
  -f(u,z) \dot{u} + \dot{z} & = 0\\
  \dot{u} + (d-1) \sqrt{-f(u,z) \dot{u}^2 + 2 \dot{u}\dot{z} + \dot{x}^2} & = 0,
\end{flalign}
which then leads to,
\begin{flalign}
  -f(u,z) u' + z' & = 0\\
  u' + (d-1) \sqrt{-f(u,z) u'^2 + 2 u' z' + 1} & = 0.
\end{flalign}
An example solution that shows how the IHRT surface behaves is shown in Fig.\ref{IHRT surface}, where the light blue dot indicates the QES.

%%%%%%%%%%%%%%%%%%%%%%%%%%%%%%%%
\section{Page Curve} \label{Page Curve}
The previous section allows for the calculation of the entanglement entropies, and so the BHRT and IHRT surfaces could be compared. In this section, we present our findings on the Page curve. However, in order for the Page curve to be calculated, the entanglement entropies should be regularized. In our setup, the regularization is provided by the pure AdS case. First, the entanglement entropy for the BHRT surface is calculated in the pure AdS along with the corresponding boundary condition imposed, similarly to how it was done in the previous section. After, the entanglement entropies of the BHRT and IHRT surfaces in the AdS-Vaidya spacetime are subtracted by the entanglement entropy calculated in the pure AdS.

\begin{figure}[h]
  \centering
  \includegraphics[scale=0.8]{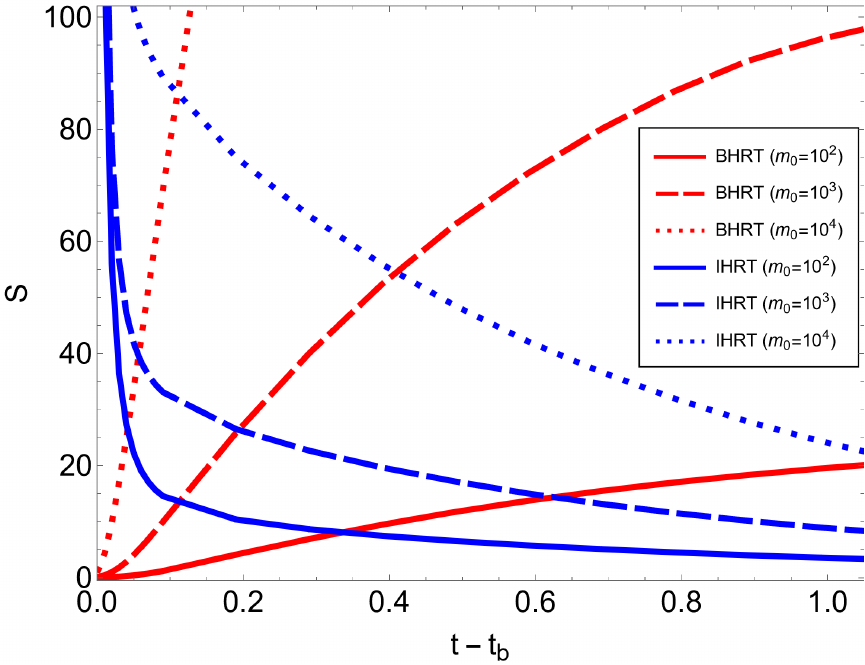}
  \caption{Phase transitions between the BHRT and IHRT in AdS-Vaidya spacetime for different initial masses $m_0 = 10^2$, $m_0 = 10^3$, and $m_0 = 10^4$. The BHRT surfaces start at zero at the initial time $t-t_b$ where $t_b$ is a cutoff. It grows monotonically in time and eventually slowly saturates. The IHRT surfaces start off from finite values and decrease monotonically 
 to zero as $t \rightarrow \infty$ which can be characterized from the evaporating black hole.}
  \label{Phase transitions}
\end{figure}

\begin{figure}[H]
  \centering
  \includegraphics[scale=0.8]{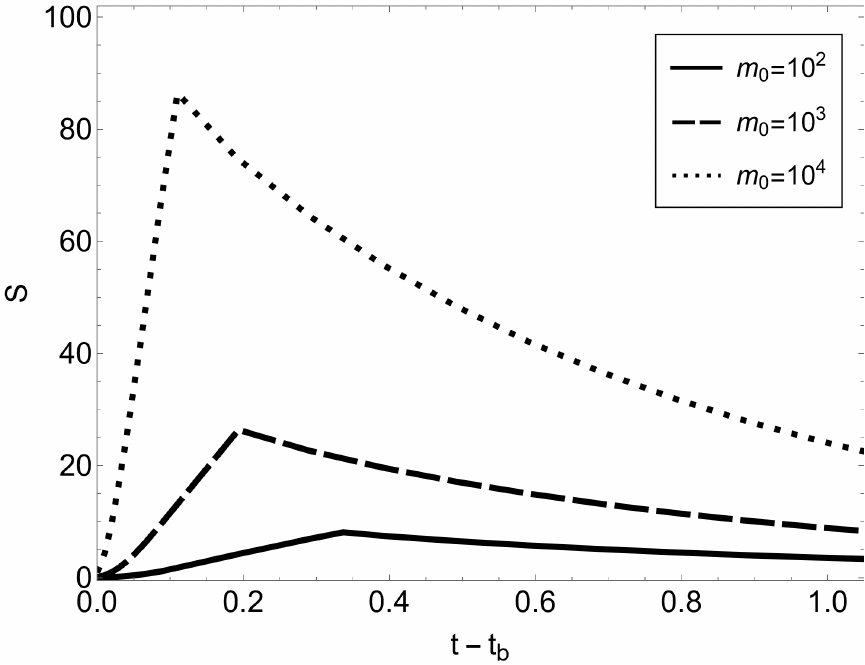}
  \caption{Page curve in AdS-Vaidya spacetime for different initial masses $m_0 = 10^2$, $m_0 = 10^3$, and $m_0 = 10^4$. The maximum value of the entanglement entropy is larger for more massive black holes, but it can be seen that the Page time occurs much earlier. This is characteristic of AdS black holes because the larger the mass the faster the rate of mass loss and implies that information can be retrieved much earlier.}
  \label{Page curve}
\end{figure}

\begin{figure}[H]
    \centering
    \includegraphics[scale=0.8]{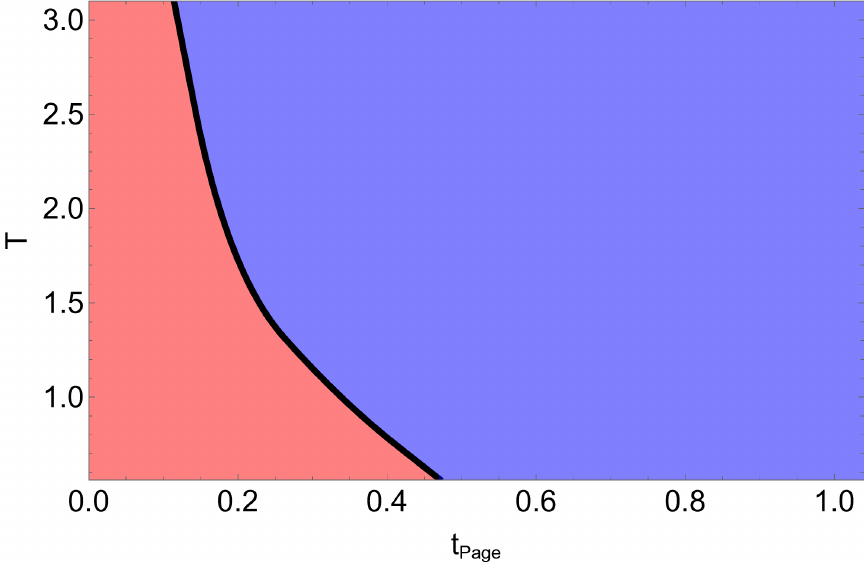}
    \caption{Phase diagram of the BHRT surface and the IHRT surface in the AdS-Vaidya spacetime. The domination of the BHRT surface is indicated by the red region, while the domination of the IHRT surface is indicated by the blue region. The phase boundary separating the BHRT surface and the IHRT surface is indicated by the black line. It can be seen that the black line increases as the Page time $t_{\rm{Page}}$ decreases, which indicates a larger temperature that corresponds to a larger AdS black hole.}
    \label{Phase diagram}
\end{figure}

The phase transition between the BHRT and IHRT surfaces for masses $m_0 = 10^2$, $m_0 = 10^3$, and $m_0 = 10^4$ are shown in Fig.\ref{Phase transitions}. The BHRT surface is depicted in red, while the IHRT surface is depicted in blue. As for the different masses, $m_0 = 10^2$ is indicated by a solid line, $m_0 = 10^3$ is indicated by a dashed line, and $m_0 = 10^4$ is indicated by a dotted line. It can be seen that the trajectories of the BHRT surfaces are initiated at zero at the initial time $t - t_b$ where $t_b$ is a cutoff, then it eventually grows monotonically. On the other hand, the IHRT surfaces start from a distinct finite value, then eventually decrease monotonically that ultimately approach zero as $t \rightarrow \infty$. The progressive decline of the IHRT surfaces can be attributed to the evaporation of the black hole.
The points of intersection between the BHRT and IHRT surfaces signal the moments of a phase transition which is from a non-island phase to an island phase where the intersection is called the Page time. Correspondingly, the Page curves for different initial masses are plotted in Fig.\ref{Page curve}. As previously, the solid, dashed, and dotted curves, respectively, represent the Page curves for the masses $m_0 = 10^2$, $m_0 = 10^3$, and $m_0 = 10^4$. An observation of the Page curve suggests that despite the maximum value of the entanglement entropy being larger for a more massive black hole, it reaches its Page time earlier. This makes sense since for AdS black holes, the larger the mass the larger the rate of mass loss as can be seen in Eq.\eqref{Mass loss}. Following the principles of entanglement wedge reconstruction, we could potentially reconstruct the information contained within the black hole at an earlier stage for a more massive black hole.
The temperature-time phase diagram for the AdS-Vaidya black hole is depicted in Fig.\ref{Phase diagram}. The red region signifies the dominance of the BHRT phase, while the IHRT phase dominates in the blue region. The black curve represents the phase boundary separating the BHRT and IHRT phases, which shows that AdS black holes with higher temperatures have an earlier Page time $t_{\rm{Page}}$. It also indirectly depicts the relation between the initial mass of a black hole to its corresponding Page time.
%%%%%%%%%%%%%%%%%%%%%%%%%%%%%%%%
\section{Conclusion} \label{Conclusion}

In this study, we have conducted a thorough investigation of the Page curve within the context of a fully time-dependent spacetime geometry, moving beyond the confines of static spacetime, partial dynamical evolution, or specific quench scenarios as explored in previous studies. Our research specifically focuses on the dynamics of an evaporating black hole modeled by the AdS-Vaidya spacetime, a scenario where the black hole undergoes gradual evaporation over an infinite temporal scale, as dictated by the Stefan-Boltzmann law, resulting in a time-variable mass.

We studied the entanglement entropy between the black hole and its emitted Hawking radiation, which is subsequently captured by a dynamically expanding CFT bath of finite size, denoted as the \textit{effective Hawking radiation region}. This innovative approach diverges from earlier methodologies, where the bath size was determined by an adjustable parameter essential for the reconstruction of the black hole interior. Our model introduces a physically motivated, naturally evolving effective Hawking radiation region, which grows in size over time without the necessity for manual parameter tuning. In addition, the reconstruction of the black hole interior is always achieved without resorting to any tuning of the parameters.

Our findings reveal a novel BCFT solution, characterized by a time-dependent METR brane. This solution signifies a departure from the conventional static ETW branes predominantly discussed in existing literature, offering a dynamic perspective where the radiation front, expanding over time, is characterized as a moving brane within the bulk spacetime. This time-dependent METR brane, aligning approximately with the light cone, introduces a significant improvement in the understanding of bulk spacetime dynamics, especially in comparison to the static ETW branes.

Moreover, the inclusion of a fully time-dependent spacetime geometry introduces a non-trivial configuration for the METR brane in the bulk spacetime. The BCFT solution, reflective of a naturally expanding effective Hawking radiation region and a time-dependent bulk geometry, emphasizes the nontriviality and novelty of our approach. The emergence of a new extremal phase, the BHRT surface, fundamentally affects the structure of the Page curve. The BHRT surface always dominates in the initial stages of evolution due to the growth of the effective Hawking radiation region. 

Our analysis extends to the interplay between the evolving spacetime geometry and the finite, time-dependent bath, focusing on the competition between the bath's growth rate and the black hole's evaporation rate. At the initial stage, given a certain rate of evaporation, the black hole emits a flux of Hawking quanta that eventually leaks into the bath. As the bath starts to grow due to the incoming Hawking quanta, at the same time, bath quanta exit the bath and enter the gravitational region, this leads to the initial increase in entanglement entropy and corresponds to the BHRT surface. At late times after the Page time, the evaporation rate is sufficiently lower than the starting rate, and the bath region grew even larger. In this scenario, the incoming Hawking flux entering the bath region cannot keep up with the bath quanta exiting the bath. This leads to the eventual decrease of the entanglement entropy and corresponds to the formation of quantum extremal islands in the gravitating region resulting in the IHRT surface.

Our research reveals that black holes of larger mass exhibit an earlier Page time, indicating a more rapid reconstruction of information from these black holes. This observation is consistent with the dynamics of mass loss for AdS black holes under our consideration, the greater the mass the faster the rate of mass loss.

Our approach advances the understanding of the intricate dynamics of black hole evaporation and provides a robust foundation in both classical and quantum gravitational theories. We propose for future work the incorporation of a thermal bath for a more realistic scenario, where the system experiences non-equilibrium situations during evaporation, offering new avenues for investigation of black hole thermodynamics.

%%%%%%%%%%%%%%%%%%%%%%%%%%%%%%%%
\section*{Acknowledgements}
We thank Xiaoning Wu for the useful discussion. This work is supported by the Ministry of Science and Technology (MOST 109-2112-M-009-005) and National Center for Theoretical Sciences, Taiwan.

%%%%%%%%%%%%%%%%%%%%%%%%%%%%%%%%
%%%\addappendix\label{Appendix}

\appendix

\renewcommand{\theequation}{\thesection.\arabic{equation}}

%%%%%%%%%%%%%%%%%%%%%%%%%%%%%%%%
\section{Eternal Black Hole} \label{Appendix Eternal Black Hole}

We review the main results of our previous work on the Page curve for an eternal black hole by using BCFT, the details of which can be found in \cite{2111.14551}. We explored the intricacies of entanglement entropy within a $(d + 1)$-dimensional two-sided eternal black hole system in a $(d + 2)$-dimensional bulk spacetime $\mathcal{M}$,
\begin{equation}
  ds_{\mathcal{M}}^2 = \frac{l^2_{AdS}}{z^2} \left[ -f(z) dt^2 + \frac{dz^2}{f(z)} + dx^2 + \sum_{i=1}^{d-1} (dx_i)^2 \right],
\end{equation}
where
\begin{equation}
  f(z) = 1 - \frac{z^{d+1}}{z_{h}^{d+1}}.
\end{equation}
The generalized entanglement entropy was computed via the utilization of the doubly holographic correspondence within a holographic BCFT framework.

We proposed the inclusion of two branes, effectively serving as the boundaries of the bulk spacetime in the BCFT framework. Firstly, the Planck brane which is characterized by $x = 0$, was introduced to embody the gravitational region of a radiating black hole within the doubly holographic arrangement. Secondly, we introduced a time-dependent METR brane\footnote{We called it time-dependent ETW brane in our previous work which is not as accurate since it can be confused with the static ETW brane mostly considered in the literature.} described by $x = t$, and is tilted away from the $t$-axis which distinguishes it from the static ETW brane that is mostly considered in the literature since the static ETW brane is parallel to the $t$-axis. This brane signifies the hypersurface of the earliest Hawking radiation which is described by a time-dependent effective Hawking radiation region $[b_L,x_L] \cup [b_R, x_R]$.

\begin{figure}[t]
  \begin{center}
    \includegraphics[width=0.9\textwidth]{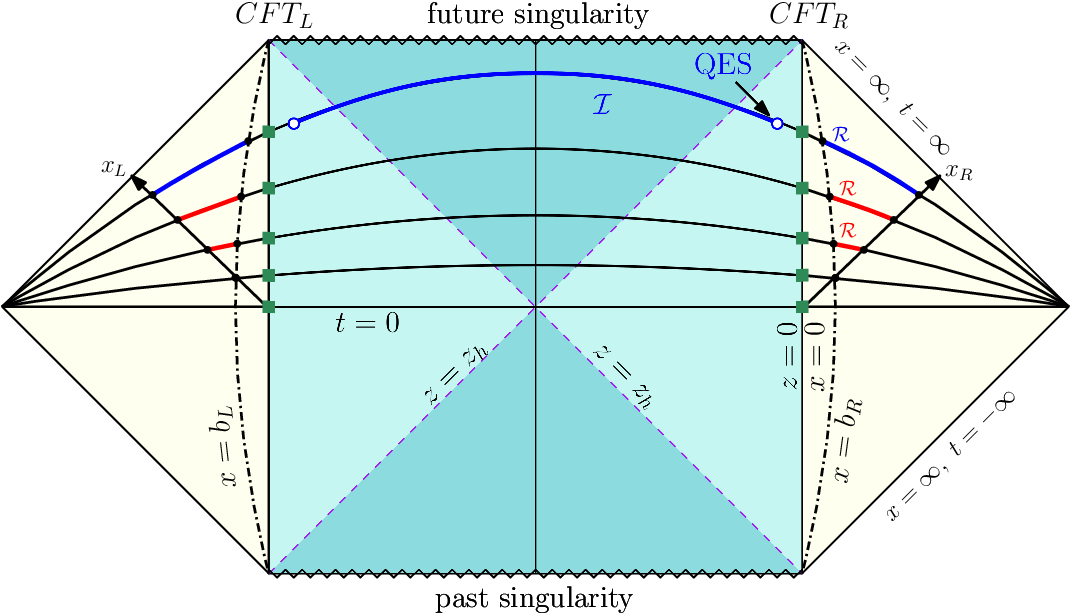}
    \caption{Penrose diagram of an eternal black hole. We highlight the effective Hawking radiation regions $\mathcal{R}$ and the late-time emergence of the island $\mathcal{I}$ along with the QES. At the initial stage, $\mathcal{R}$ is colored in red to indicate the early dominance of the BRT surface, while at late times $\mathcal{R}$ is colored in blue to indicate the late time dominance of the IRT surface.}
    \label{Penrose_Eternal}
  \end{center}
\end{figure}

We studied various RT surfaces associated with holographic entanglement entropy. One such surface is the Boundary RT (BRT) surface which intersects the time-dependent METR brane and exhibits temporal growth due to the time-dependent nature of the METR brane. On the other hand, the Island RT (IRT) surface which intersects the Planck brane and provides support to an island remains time-independent. The contribution of the BRT surface at early times and the eventual emergence of the island bounded by the QES at late times represented by the IRT surface is shown in the Penrose diagram in Fig.\ref{Penrose_Eternal}. Furthermore, we conducted an analysis of the phase transitions between the aforementioned RT surfaces. The point of transition, also known as the Page time, was determined by equating their corresponding entanglement entropies. Consequently, the phase diagram of the entanglement entropy was established.

It is noteworthy that in actual black hole evaporation scenarios, the temperature varies over time. Therefore, it is essential to consider the temperature dependence of the phase transition and to determine the corresponding Page curve. The present study focuses on addressing these temperature-dependent dynamics in the context of a time-dependent black hole spacetime.

%%%%%%%%%%%%%%%%%%%%%%%%%%%%%%%%
\section{ANEC}\label{Appendix ANEC}

In examining the ANEC, we follow closely the approach taken in \cite{2111.05151}. We consider a $(d+2)$-dimensional asymptotically AdS-Vaidya evaporating black hole where the negative energy flux allows the black hole mass to decrease. In order to investigate the ANEC along the event horizon, it is suitable to use the ingoing AdS-Vaidya metric,
\begin{equation} \label{Ingoing AdS Vaidya Metric}
  ds^2_{\mathcal{M}} = \frac{l^2_{AdS}}{z^2} \left[ -f(v,z) dv^2 - 2dvdz + \sum_{i=1}^{d} (dx_i)^2  \right],
\end{equation}
with a mass function given by,
\begin{equation}
  m(v) = m_0 e^{-\sigma v}, \qquad f(v,z) = 1 - m(v) z^{d+1},
\end{equation}
where $m_0$ is the initial mass and the area decreases starting from $v=v_h$ until $v=v_f$ for which the black hole completely evaporates. The value $v=v_h$ corresponds to the time at the boundary CFT where the radiation starts to leak into the thermal bath. In the case of an AdS black hole, it takes an infinite amount of time to evaporate, i.e. $m \rightarrow 0$ as $v_f \rightarrow \infty$. We note that in our setup, we consider the case where $d=3$ so that we have a $5$-dimensional spacetime.

The expression for the radial null geodesic along the event horizon is given by,
\begin{equation}\label{radial null geodesic}
  \frac{dz}{dv} = -\frac{1}{2} \left( 1 - m(v) z^4 \right).
\end{equation}

It is important to note that during the evaporation, the apparent horizon given by $z_{AH} = m(v)^{-1/4}$ is located outside the event horizon, and it is an upper bound of the event horizon. Even though the evaporation model considered here is quasi-static, i.e., extremely slow evaporation, there should be an abrupt shift in the geometry near the end of the evaporation for which it is conceivable to presume that the event horizon may be much smaller than the apparent horizon. In the $z$ coordinate, a smaller event horizon corresponds to a large $z_{EH}$, for which Eq.\eqref{radial null geodesic} can be approximated in the limit near the end of the evaporation,
\begin{equation}
  \frac{dz}{dv} \approx \frac{m(v) z^4}{2},
\end{equation}
which when integrated gives us the event horizon $z_{\rm{EH}}$ as a function of $v$,
\begin{equation}\label{zeh approx}
  z_{\rm{EH}} = \left( \frac{2 \sigma }{3 m_0 e^{-\sigma v}} \right)^{1/3}.
\end{equation}
The radial outgoing null vector can be written as,
\begin{equation}
  \ell^\mu = (\partial_v)^\mu - \frac{1}{2} \left( 1 - m(v) z^4 \right) (\partial_z)^\mu.
\end{equation}
Along the event horizon, the null geodesic generator $\alpha$ parameterized by an affine parameter satisfies,
\begin{equation}
  \mathcal{\xi}^\mu = h(v) \ell^\mu,
\end{equation}
where the null geodesic affine parameter is denoted by $V$ corresponding to the tangent vector $\mathcal{\xi}^\mu$, and $h(v)$ satisfies the differential equation,
\begin{equation}
  \frac{dh}{dv} = -h \left( \frac{1 + m(v) z^4}{z} \right).
\end{equation}
As $v \rightarrow v_f$, $h$ can be approximated as,
\begin{equation}\label{h approx}
  h = \frac{dv}{dV} \propto e^{\frac{-2 \sigma v}{3}}.
\end{equation}
Next, we construct the 6-dimensional bulk written in Fefferman-Graham coordinates,

\begin{equation}\label{Fefferman-Graham}
  \begin{aligned}
     g_{ab} dx^a dx^b &= \frac{L^2_{AdS}}{\rho^2} \left( d\rho^2 + g_{\mu\nu}(x,\rho) dx^\mu dx^\nu \right) = \frac{L^2_{AdS}}{\rho^2} \hat{g}_{ab} dx^a dx^b,\\
     g_{\mu\nu}(x,\rho) &= g_{(0) \mu\nu}(x) + \rho^2 g_{(2) \mu\nu}(x) \ldots + \rho^d g_{(d) \mu\nu}(x) + h_{(d) \mu\nu} \rho^d \rm{ln} \rho^2 + \ldots,
  \end{aligned}
\end{equation}

where $g_{(0) \mu\nu}$ is the boundary metric. The logarithmic term appears only for even boundary dimension $d$, and $g_{(2k+1) \mu \nu} = 0$ for all integers $k$ satisfying $0 \leq 2k+1 < d$. Note that as an abuse of notation, here and in the following, we use $d$ to indicate the dimensionality of the whole boundary spacetime, i.e., $d=5$. The coefficients $g_{(2) \mu\nu}$ and $g_{(4) \mu\nu}$ of the subleading terms are then given by \cite{0002230}\footnote{The convention for the curvature used here is so that the curvature of AdS comes out negative.},
\begin{eqnarray}
    g_{(2) \mu\nu} &=& -\frac{1}{d-2} \left( R_{\mu\nu} - \frac{1}{2 (d-1)} R g_{(0) \mu\nu}\right),\\
     g_{(4) \mu\nu} &=& \frac{1}{d-4} \Bigg( \frac{1}{8 (d-1)} \nabla_\mu \nabla_\nu R - \frac{1}{4 (d-2)} \nabla^\alpha \nabla_\alpha R_{\mu \nu} \nonumber\\
    & & + \frac{1}{8 (d-1) (d-2)} g_{(0) \mu \nu} \nabla^\alpha \nabla_\alpha R - \frac{1}{2 (d-2)} R^{\alpha \beta} R_{\mu \alpha \nu \beta} + \frac{d-4}{2 (d-2)^2} R^{\; \alpha}_{\mu} R_{\alpha \nu}\\
    & & + \frac{1}{(d-1) (d-2)^2} R R_{\mu \nu} + \frac{1}{4 (d-2)^2} R^{\alpha \beta} R_{\alpha \beta} g_{(0) \mu \nu} - \frac{3 d}{16 (d-1)^2 (d-2)^2} R^2 g_{(0) \mu \nu} \Bigg),\nonumber
\end{eqnarray}
where $R$ is the Ricci scalar and $R_{\mu \nu}$ is the Ricci tensor of the boundary metric $g_{(0) \mu\nu}$.

The coefficient $g_{(d) \mu\nu}$ is correlated to the renormalized stress-energy tensor on the boundary,
\begin{equation}
    \expval{T_{\mu\nu}} = \frac{dl^{d-1}}{16 \pi G} g_{(d) \mu\nu} + X_{\mu \nu},
\end{equation}
where $X_{\mu \nu}$ corresponds to the gravitational conformal anomaly which is non-vanishing only for boundaries of even dimension.

In applying the no-bulk-shortcut property, it is convenient to express the boundary metric in double-null coordinates,
\begin{equation}
    g_{(0) \mu \nu} dx^\mu dx^\nu = -e^{f(U,V)} dU dV + r^2(U,V) \sum_{i=1}^3 dx_i^2,
\end{equation}
where we can apply the coordinate change $r = 1/z$ and absorb the factor appearing in $dUdV$ to $f(U,V)$, and then solving the equation of motion from there. It is important to note that along the event horizon, we can set $U=0$ and $f(0,V) = 0$ without loss of generality.

Given a boundary null geodesic $\alpha$ parameterized by the affine parameter $V$ with the tangent vector $\mathcal{\xi}^\mu$, we denote by $V_i$ and $V_f$ the values of the two points that $\alpha$ passes through. These two points may indicate the start and end of the whole process, i.e., empty spacetime up until the black hole completely evaporates. Near the boundary null geodesic $\alpha$, we can now consider a bulk causal curve $\beta$ with the tangent vector $\mathcal{\xi}^a$,
\begin{equation}\label{Bulk Tangent}
    \mathcal{\xi}^a = \left( \mathcal{\xi}^{\rho}, \mathcal{\xi}^U, \mathcal{\xi}^V, \mathcal{\xi}^{x_1}, \mathcal{\xi}^{x_2}, \mathcal{\xi}^{x_3} \right) = \left( \frac{d\rho}{dV}, \frac{dU}{dV}, 1, 0, 0, 0 \right),
\end{equation}
where it is adequate to suppose $\mathcal{\xi}^{x_1} = \mathcal{\xi}^{x_2} = \mathcal{\xi}^{x_3} = 0$ from the boundary metric considered.

An expansion in the small parameter $\epsilon$ can be considered,
\begin{eqnarray}
     \rho &=& \rho_1 \epsilon + \rho_2 \epsilon^2 + \ldots,\label{Expansion1}\\
     \frac{dU}{dV} &=& \frac{dU_2}{dV} \epsilon^2 + \frac{dU_3}{dV} \epsilon^3 + \ldots.
     \label{Expansion2}
\end{eqnarray}
Using the constraint for the no-bulk-shortcut $\hat{g}_{ab} K^a K^b \leq 0$ and substituting Eq.\eqref{Bulk Tangent}, Eq.\eqref{Expansion1}, and Eq.\eqref{Expansion2} into Eq.\eqref{Fefferman-Graham} we get,
\begin{equation}\label{orders}
\begin{split}
    \hat{g}_{ab} K^a K^b & = \epsilon^2 \left( \dot{\rho}_1^2 + \rho_1^2 g_{(2)VV} (0,V) -\frac{dU_2}{dV} \right)\\ 
    & + \epsilon^3 \left( 2 \dot{\rho}_1 \dot{\rho}_2 + 2 \rho_1 \rho_2 g_{(2)VV} (0,V) -\frac{dU_3}{dV} \right)\\
    & + \epsilon^4 \Bigg( \dot{\rho}_2^2 + 2 \dot{\rho}_1 \dot{\rho}_3 + \rho_2^2 g_{(2)VV} (0,V) + 2 \rho_1 \rho_3 g_{(2)VV} (0,V)\\
    & \qquad + \rho_1^4 g_{(4)VV} (0,V) + 2 \rho_1^2 g_{(2)UV} (0,V) \frac{dU_2}{dV} - \frac{dU_4}{dV}\\
    & \qquad + \rho_1^2 \partial_U (g_{(2)VV}) (0,V) U_2 - (\partial_U f) U_2 \frac{dU_2}{dV} \Bigg)\\
    & + \epsilon^5 \Bigg( 2 \dot{\rho}_2 \dot{\rho}_3 + 2 \dot{\rho}_1 \dot{\rho}_4 + 2 \rho_2 \rho_3 g_{(2)VV} (0,V) + 2 \rho_1 \rho_4 g_{(2)VV} (0,V)\\
    & \qquad + 4 \rho_1^3 \rho_2 g_{(4)VV} (0,V) + \rho_1^5 g_{(5)VV} (0,V) + 2 \rho_1^2 g_{(2)UV} (0,V) \frac{dU_3}{dV}\\
    & \qquad + 4 \rho_1 \rho_2 g_{(2)UV} (0,V) \frac{dU_2}{dV} - \frac{dU_5}{dV} + \rho_1^2 \partial_U (g_{(2)VV}) (0,V) U_3\\
    & \qquad + 2 \rho_1 \rho_2 \partial_U (g_{(2)VV}) (0,V) U_2 - (\partial_U f) U_2 \frac{dU_3}{dV} - (\partial_U f) U_3 \frac{dU_2}{dV} \Bigg) \leq 0,\\
\end{split}
\end{equation}
where the derivative with respect to $V$ is denoted by a dot.

We calculate the inequality concerning the variation of $U$ order by order and note that the equality is satisfied by the null curve, 
\begin{equation}
    \Delta U_2 = \int_{V_i}^{V_f} \frac{dU_2}{dV} dV \geq \int_{V_i}^{V_f} \left( \dot{\rho}_1^2 + \rho_1^2 g_{(2)VV} (0,V) \right) dV,
\end{equation}
where doing the variation on the right side of the inequality leads to,
\begin{equation}
    \ddot{\rho}_1 = g_{(2)VV} (0, V) \rho_1 = \frac{\ddot{r}(0, V)}{r(0, V)} \rho_1,
\end{equation}
which has a solution satisfying the boundary condition $\rho_1 (V_i) = \rho_1 (V_f) = 0$,
\begin{equation}\label{Z1 approx}
    \rho_1 = c r(0,V),
\end{equation}
where $c$ is a positive constant and one can easily find that $U_2 = \rho_1 \dot{\rho}_1$. The limit of $U_2$ can also be taken,
\begin{equation}
    \lim_{V \rightarrow V_f} \rho_1(V) \dot{\rho}_1(V) = \lim_{v \rightarrow v_f} \rho_1(V) h(v) \frac{d\rho_1}{dv} \rightarrow 0,
\end{equation}
where Eq.\eqref{zeh approx}, Eq.\eqref{h approx}, and Eq.\eqref{Z1 approx} were used. This ensures that the bulk causal curve stays in the neighborhood of the boundary null geodesic $\alpha$ in the range $[V_i,V_f]$ for small $\epsilon$. Since the right side of the inequality vanishes, it implies that when $\beta$ is a null curve there should be no time delay between $\alpha$ and $\beta$ at $O(\epsilon^2)$.

A similar scenario happens for $O(\epsilon^3)$, where $\rho_2$ satisfies a similar solution as in Eq.\eqref{Z1 approx} while $U_3 = 2\rho_2 \dot{\rho}_2$. For $O(\epsilon^4)$, the first four terms can be shown to vanish by variation where $\rho_3$ satisfies a similar solution as in Eq.\eqref{Z1 approx}. All that is left is to show that the leftover terms are nonnegative since $\Delta U_4$ must be nonnegative which is what the no-bulk-shortcut property requires,
\begin{equation}
\begin{split}
\Delta U_4 \geq & \int_{V_i}^{V_f} \Bigg( \rho_1^4 g_{(4)VV} (0,V) + 2 \rho_1^2 g_{(2)UV} (0,V) \frac{dU_2}{dV}\\
& + \rho_1^2 \partial_U (g_{(2)VV}) (0,V) U_2 - (\partial_U f) U_2 \frac{dU_2}{dV} \Bigg) dV.
\end{split}
\end{equation}
The leftover terms are calculated as follows,
\begin{equation}
\begin{split}
2 \int_{V_i}^{V_f} \rho_1^2 g_{(2)UV} \frac{dU_2}{dV} dV & = 2 \rho_1^2 g_{(2)UV} U_2 \Big| - 2 \int_{V_i}^{V_f} \partial_V \left( \rho_1^2 g_{(2)UV} \right) U_2 dV\\ 
& = -2 c^4 \int_{V_i}^{V_f} \partial_V \left( -\frac{r^2}{4}  \right) r \dot{r} dV = c^4 \int_{V_i}^{V_f} r^2 \dot{r}^2 dV,
\end{split}
\end{equation}

\begin{equation}
\begin{split}
\int_{V_i}^{V_f} \rho_1^2 \partial_U (g_{(2)VV}) U_2 dV & = \int_{V_i}^{V_f} r^3 \dot{r} \partial_U (g_{(2)VV}) dV = \int_{V_i}^{V_f} r^3 \dot{r} \partial_U \left( \frac{\ddot{r} - \dot{f} \dot{r}}{r}\right) dV\\ 
& = c^4 \int_{V_i}^{V_f} (r \dot{r} r' \ddot{r} + r^3 \ddot{r} - 2 \dot{r}^3 r' + r^2 \dot{r}^2) dV,
\end{split}
\end{equation}

\begin{equation}
\begin{split}
 - \int_{V_i}^{V_f} (\partial_U f) U_2 \frac{dU_2}{dV} dV & = -f' U_2^2 \Big| + \int_{V_i}^{V_f} \dot{f}' U_2^2 dV + \int_{V_i}^{V_f} f' U_2 \frac{dU_2}{dV} dV\\ 
&= \frac{1}{2} \int_{V_i}^{V_f} \dot{f}' U_2^2 dV = \frac{1}{2} c^4 \int_{V_i}^{V_f} r^2 \dot{r}^2 \dot{f}' dV\\
 &= c^4 \int_{V_i}^{V_f} \left( 3 \dot{r}^3 r' + \frac{1}{2} r^2 \dot{r}^2 \right) dV,
\end{split}
\end{equation}

\begin{equation}
\begin{split}
\int_{V_i}^{V_f} \rho_1^4 g_{(4)VV} (0,V) dV & = \frac{1}{8} c^4 \int_{V_i}^{V_f} \Big( 12 r \dot{r} r' \ddot{r} + 2 r^2 r' \dddot{r} - 2 r^2 \dot{r}' \ddot{r} - 10 r^2 \dot{r} \ddot{r}' + r^4 \dddot{f}'\\
& \qquad \qquad + 8 r^3 \dot{r} \ddot{f}' - 2 r^3 \dddot{r}' + 4 \dot{r}^3 r' - 4 r \dot{r}^2 \dot{r}' + 12 r^2 \dot{r}^2 \dot{f}' \Big) dV \\ 
& = \frac{1}{8} c^4 \int_{V_i}^{V_f} (-16 r \dot{r} r' \ddot{r} - 6 r^3 \ddot{r} - 4 \dot{r}^3 r' - 10 r^2 \dot{r}^2) dV\\
& = c^4 \int_{V_i}^{V_f} (-2 r \dot{r} r' \ddot{r} - \frac{3}{4} r^3 \ddot{r} - \frac{1}{2} \dot{r}^3 r' - \frac{5}{4} r^2 \dot{r}^2) dV,
\end{split}
\end{equation}
where applying the transformation to the Ricci tensor $R_{\mu \nu}$, while taking the component $R_{U V}$ and the transverse component $R_{x x}$ we obtain the expressions,
\begin{eqnarray}
    \dot{r}' &=& -\frac{1}{r} ( 2 \dot{r} r' + r^2 ),\\
    \dot{f}' &=& -\frac{1}{r} ( 3 \dot{r}' + 2 r ),
\end{eqnarray}
where the derivative with respect to $U$ is denoted by a prime. These expressions are used in the leftover terms. In addition, the boundary terms obtained in the intermediate step can also be shown to vanish. 

For $O(\epsilon^5)$, the first four terms can be similarly shown to vanish for the corresponding solution established above so that we are left with,
\begin{equation}
\begin{split}
\Delta U_5 \geq & \int_{V_i}^{V_f} \Bigg( \rho_1^5 g_{(5)VV} (0,V)  + 4 \rho_1^3 \rho_2 g_{(4)VV} (0,V) + 2 \rho_1^2 g_{(2)UV} (0,V) \frac{dU_3}{dV}\\
& \qquad + 4 \rho_1 \rho_2 g_{(2)UV} (0,V) \frac{dU_2}{dV} + \rho_1^2 \partial_U (g_{(2)VV}) (0,V) U_3\\
& \qquad + 2 \rho_1 \rho_2 \partial_U (g_{(2)VV}) (0,V) U_2 - (\partial_U f) U_2 \frac{dU_3}{dV} - (\partial_U f) U_3 \frac{dU_2}{dV} \Bigg) dV.
\end{split}
\end{equation}
Applying the no-bulk-shortcut property requires that $\Delta U_5$ be nonnegative so that we have,
\begin{equation}\label{ANEC}
 \int_{V_i}^{V_f} \rho_1^5 g_{(5)VV} (0,V) dV \geq \mathcal{I},
\end{equation}
\begin{equation}\label{Inonnegative}
\begin{split}
 \mathcal{I} = & - \int_{V_i}^{V_f} \Bigg( 4 \rho_1^3 \rho_2 g_{(4)VV} (0,V) + 2 \rho_1^2 g_{(2)UV} (0,V) \frac{dU_3}{dV}\\
& + 4 \rho_1 \rho_2 g_{(2)UV} (0,V) \frac{dU_2}{dV} + \rho_1^2 \partial_U (g_{(2)VV}) (0,V) U_3\\
 & + 2 \rho_1 \rho_2 \partial_U (g_{(2)VV}) (0,V) U_2 - (\partial_U f) U_2 \frac{dU_3}{dV} - (\partial_U f) U_3 \frac{dU_2}{dV} \Bigg) dV,
\end{split}
\end{equation}
where the inequality that we need for ANEC is given by Eq.\eqref{ANEC} and all we have to do now is to show that Eq.\eqref{Inonnegative} is nonnegative to prove that ANEC holds along with a weight factor.

Since for $i=1, \ldots, 4$, all $\rho_i$ have a similar form, while $U_2 = \rho_1 \dot{\rho}_1$ and $U_3 = 2\rho_2 \dot{\rho}_2$, it can be observed that all the terms have already been calculated in $O(\epsilon^4)$ so that $\mathcal{I}$ becomes,
\begin{comment}
\begin{equation}
\begin{split}
& \int_{V_i}^{V_f} \left( 2 \rho_1^2 g_{(2)UV} (0,V) \frac{dU_3}{dV} + 4 \rho_1 \rho_2 g_{(2)UV} (0,V) \frac{dU_2}{dV} \right) dV\\
& = c^4 \int_{V_i}^{V_f} 4 r^2 \dot{r}^2 dV,
\end{split}
\end{equation}

\begin{equation}
\begin{split}
& \int_{V_i}^{V_f} \left( \rho_1^2 \partial_U (g_{(2)VV}) U_3 + 2 \rho_1 \rho_2 \partial_U (g_{(2)VV}) U_2 \right) dV\\
& = c^4 \int_{V_i}^{V_f} (4 r \dot{r} r' \ddot{r} + 4 r^3 \ddot{r} - 8 \dot{r}^3 r' + 4 r^2 \dot{r}^2) dV,
\end{split}
\end{equation}

\begin{equation}
\begin{split}
& \int_{V_i}^{V_f} \left( - (\partial_U f) U_2 \frac{dU_3}{dV} - (\partial_U f) U_3 \frac{dU_2}{dV} \right) dV\\
& = c^4 \int_{V_i}^{V_f} ( 12 \dot{r}^3 r' + 2 r^2 \dot{r}^2) dV,
\end{split}
\end{equation}

\begin{equation}
\begin{split}
& \int_{V_i}^{V_f} \left( 4 \rho_1^3 \rho_2 g_{(4)VV} (0,V) \right) dV\\
& = c^4 \int_{V_i}^{V_f} (-8 r \dot{r} r' \ddot{r} - 3 r^3 \ddot{r} - \dot{r}^3 r' - 5 r^2 \dot{r}^2) dV,
\end{split}
\end{equation}    
\end{comment}
\begin{equation}\label{Iequation}
    \mathcal{I} = c^4 \int_{V_i}^{V_f} \left( 4 r \dot{r} r' \ddot{r} - r^3 \ddot{r} - 2 \dot{r}^3 r' - 5 r^2 \dot{r}^2 \right) dV.
\end{equation}
Writing the variables in terms of $z$ and noting that $\dot{r} = h \frac{dr}{dv}$, while $r' = 1/(-2 h)$ with $\partial_U = -\frac{1}{2 h} \partial_r$, and $f(v,z) = 1 - m(v) z^4$,
\begin{eqnarray}
    \frac{dr}{dv} &=& \frac{1}{2 z^2} f(v,z),\\
    \frac{d^2r}{dv^2} &=& -\frac{z^2}{2} \frac{dm(v)}{dv} + m(v) z f(v,z) + \frac{1}{2 z^3} f(v,z)^2
\end{eqnarray}
We can now rewrite $\mathcal{I}$ as,
\begin{equation}\label{Ifinal}
\begin{split}
\mathcal{I} = & c^4 \int_{v_i}^{v_f} h \Bigg[ \left( -\frac{1}{z^3} f(v,z) -\frac{1}{z^3} \right) \Bigg( -\frac{z^2}{2} \frac{dm(v)}{dv} + m(v) z f(v,z)\\
& \qquad + \frac{1}{2 z^3} f(v,z)^2 - \frac{1}{2 z^3} f(v,z) (1 + m(v) z^4) \Bigg)\\
& \qquad + \frac{1}{8 z^6} f(v,z)^3 - \frac{5}{4 z^6} f(v,z)^2 \Bigg] dv. 
\end{split}
\end{equation}
To study this, we can separate the whole region $v_i \leq v \leq v_f$ into phases,

Phase (I) $v_i \leq v \leq v_0 - \delta$: pure AdS spacetime,

Phase (II) $v_0 - \delta \leq v \leq v_0 + \delta$: collapse of null shell,

Phase (III) $v_0 + \delta \leq v \leq v_h$: AdS-Schwarzschild spacetime,

Phase (IV) $v_h \leq v < v_f$: black hole evaporation.

In this scenario, for the AdS case the mass is zero so Eq.\eqref{Ifinal} vanishes trivially. We assume that the duration $\delta$ is extremely small such that $\delta \ll | v_f - v_0|$ so that the collapse occurs rapidly and the mass grows quickly until it settles to the AdS-Schwarzschild case with mass $m_0$. During the collapse, the dominant term is the product $(-1/z^3)(-z^2/2)(dm/dv)$ while the rest are negligible. As for the AdS-Schwarzschild case, the mass is constant and so all terms result to zero.

For the evaporation, we consider the case where it is extremely slow, i.e., $m_0 \ll v_f$. Noting that the event horizon is located inside the apparent horizon, we can set the location of the event horizon as,
\begin{equation}
    z = \frac{1}{m^{1/4} (1-\eta)^{1/4}}
\end{equation}
where $|\eta| \sim |dm/dv| \ll 1$. Using this approximation, the only term that contributes is also the product $(-1/z^3)(-z^2/2)(dm/dv)$. So we have,
\begin{equation}
    \frac{h}{2} \int_{v_0 - \delta}^{v_0 + \delta} m^{1/4} \frac{dm}{dv} dv + \frac{h}{2} \int_{v_h}^{v_f} m^{1/4} \frac{dm}{dv} dv = \frac{2}{5} h \left( m^{5/4} \big|_0^{m_0} + m^{5/4} \big|_{m_0}^ 0 \right) = 0.
\end{equation}
Thus we have,
\begin{equation}
    \int_{V_i}^{V_f} \rho_1^5 g_{(5)VV} (0,V) dV = \int_{V_i}^{V_f} \rho_1^5 \expval{T_{\mu \nu}} dV \geq 0.
\end{equation}
This shows that ANEC with a corresponding weight factor is satisfied.

%%%%%%%%%%%%%%%%%%%%%%%%%%%%%%%%
\bibliographystyle{unsrt}
\bibliography{Page_Curve_of_AdS_Vaidya}

%%%%%%%%%%%%%%%%%%%%%%%%%%%%%%%%

\end{document}